\documentclass[conference]{IEEEtran}
\pdfoutput=1
\IEEEoverridecommandlockouts
\usepackage{cite}
\usepackage{amsmath,amssymb,amsfonts}
\usepackage{algorithmic}
\usepackage{caption}
\usepackage{multirow}
\usepackage{algorithm}
\usepackage{graphicx}
\usepackage{siunitx}
\usepackage{dblfloatfix}
\usepackage{textcomp}
\usepackage{xcolor}
\usepackage{booktabs}
\usepackage[tight,footnotesize]{subfigure}
\usepackage{epsfig}
\def\BibTeX{{\rm B\kern-.05em{\sc i\kern-.025em b}\kern-.08em
    T\kern-.1667em\lower.7ex\hbox{E}\kern-.125emX}}

\mathchardef\mhyphen="2D

\begin{document}

% original title
\title{FedSZ: Leveraging Error-Bounded Lossy Compression for Federated Learning Communications\vspace{-10mm}}
% arXiv title
%\title{Efficient Communication in Federated Learning Using Floating-Point Lossy Compression\vspace{-10mm}}

\author{\IEEEauthorblockN{Grant Wilkins\IEEEauthorrefmark{1}\IEEEauthorrefmark{2},
Sheng Di\IEEEauthorrefmark{1}
Jon C. Calhoun\IEEEauthorrefmark{3},
Zilinghan Li\IEEEauthorrefmark{2}\IEEEauthorrefmark{4},
Kibaek Kim\IEEEauthorrefmark{1},
Robert Underwood\IEEEauthorrefmark{1},
\\
Richard Mortier\IEEEauthorrefmark{2},
Franck Cappello\IEEEauthorrefmark{1}}
\IEEEauthorblockA{\IEEEauthorrefmark{1}Argonne National Laboratory, Lemont, IL, USA}
\IEEEauthorblockA{\IEEEauthorrefmark{2}University of Cambridge, Cambridge, UK}
\IEEEauthorblockA{\IEEEauthorrefmark{3}Clemson University, Clemson, SC, USA}
\IEEEauthorblockA{\IEEEauthorrefmark{4}University of Illinois at Urbana-Champaign, Urbana, IL, USA}

Emails: \{gfw27,rmm1002\}@cam.ac.uk, \{sdi1,kimk,runderwood,cappello\}@anl.gov, jonccal@clemson.edu, zl52@illinois.edu
}

\maketitle

\begin{abstract}
% ORIGINAL ABSTRACT
With the promise of federated learning (FL) to allow for geographically-distributed and highly personalized services, the efficient exchange of model updates between clients and servers becomes crucial. FL, though decentralized, often faces communication bottlenecks, especially in resource-constrained scenarios. Existing data compression techniques like gradient sparsification, quantization, and pruning offer some solutions, but may compromise model performance or necessitate expensive retraining. In this paper, we introduce \textsc{FedSZ}, a specialized lossy-compression algorithm designed to minimize the size of client model updates in FL. \textsc{FedSZ} incorporates a comprehensive compression pipeline featuring data partitioning, lossy and lossless compression of model parameters and metadata, and serialization. We evaluate \textsc{FedSZ} using a suite of error-bounded lossy compressors, ultimately finding \texttt{\texttt{SZ2}} to be the most effective across various model architectures and datasets including AlexNet, MobileNetV2, ResNet50, CIFAR-10, Caltech101, and Fashion-MNIST. Our study reveals that a relative error bound $10^{-2}$ achieves an optimal tradeoff, compressing model states between $5.55\mhyphen12.61\times$ while maintaining inference accuracy within $<0.5\%$ of uncompressed results. Additionally, the runtime overhead of \textsc{FedSZ} is $<4.7\%$ or between of the wall-clock communication-round time, a worthwhile trade-off for reducing network transfer times by an order of magnitude for networks bandwidths $<500\si{Mbps}$. Intriguingly, we also find that the error introduced by \textsc{FedSZ} could potentially serve as a source of differentially private noise, opening up new avenues for privacy-preserving FL.
\end{abstract}

\begin{IEEEkeywords}
lossy compression, federated learning, Internet of Things (IoT)
\end{IEEEkeywords}

\section{Introduction}
Federated learning (FL), a decentralized machine learning (ML) paradigm, has found applications in a plethora of fields, including image classification and generative text models on mobile phones, real-time decision-making on edge devices, large-scale anomaly detection, and personalization in healthcare~\cite{federated-applications}. However, FL's broad adoption and effectiveness face challenges due to the high complexity of models, substantial computational demands and ever-increasing scale. Modern FL systems often require processing and training models, each with billions of parameters, while communicating with thousands of distributed clients, bringing a heavy computational and communication overhead. 

In FL, client-server communication can become a significant bottleneck to achieving scalability of number of clients and server-side robustness~\cite{scale, scale2, scale3}. For instance, an autonomous vehicle can generate up to 1 GB/s of sensor or image data for on-device training and validation~\cite{autonomous-data-size}. If a 10GB client update is sent to a server via a mobile network with a bandwidth of $10\si{Mbps}$, it would take approximately 150 minutes to transmit. With several other computational tasks happening and an autonomous vehicle being battery-constrained, this is a significant amount of effort spent on communication alone. As a result, strategies to cut data-related communication challenges are necessary to enhance the scalability and robustness of FL~\cite{challenges}.

Error-bounded lossy compression (EBLC)~\cite{sz3-1, sz3-2, zfp} is widely used to significantly reduce large volumes of data generated by high-performance computing (HPC) simulations, but how EBLC can be used to mitigate the communication cost for FL significantly is still an open question. Moreover, introducing EBLC produces many new challenges for FL: (1) The FL environment fundamentally differs from HPC. In the area of HPC, the compute or storage nodes are generally under centralized management in a supercomputer, where the related resources are relatively stable. However, in FL, the distributed clients could be any device (e.g., Raspberry Pi) on a wide area network (WAN), projecting a fairly non-deterministic environment with diverse communication bandwidths and node compute power. So, a lossy compression method has to consider data fidelity, network latency, and device processing power. (2) Scientific datasets are generally much more compressible than the parameter datasets generated by FL clients. The key reason is that the scientific datasets generally correspond to simulating or capturing a physical, chemical or biological phenomenon. In comparison, the FL parameter data are often irregular/spiky, to be shown in Section \ref{subsec:bad-data}.

In this paper, we address several key questions to confirm our hypothesis that \textit{EBLC significantly cuts communication costs and overall wall-clock time while preserving FL model accuracy very well.} 
\begin{enumerate}
    \item How can we integrate EBLC into the FL communication pipeline?
    \item Given the uniqueness and complexity of model parameters in FL, which EBLC algorithm yields the best trade-off between size reduction and retaining model data fidelity?
    \item Lossy compression, while reducing data size, introduces both computational overhead and data error. Is the runtime and error cost incurred by these factors outweighed by the benefits of reduced communication time?
\end{enumerate}

In this study, we present a novel optimized, practical EBLC heuristic, \textsc{FedSZ}, designed to compress FL client model updates and significantly reduce client-server communication costs. In particular, we make the following contributions:

\begin{itemize}
    \item We share \textsc{FedSZ} as an open-source tool integrated with the Advanced Privacy-Preserving Federated Learning (APPFL) package~\cite{appfl}, which effectively compresses any PyTorch compatible FL model with EBLC and lossless compression. Our study is the first attempt to reduce communication overhead for FL through EBLC. 
    \item We comprehensively characterize both EBLC and lossless compression algorithms to discover which combination most effectively reduces the size of FL updates while maintaining a low runtime and minimal impact on model capabilities. We find that \texttt{SZ2} and \texttt{blosc-lz} are the best compressors for low-runtime, data-reduction, and accuracy-preservation from our test suite.
    \item We evaluate \textsc{FedSZ}'s communication savings by performing rounds of FL on a cluster and simulating different network bandwidths. Experimental results confirm that we achieve (i) $5.55$\textendash$12.61\times$ space savings in client updates while keeping uncompressed inference accuracy, (ii) 13.26$\times$ communication time savings for a $10\si{Mbps}$ bandwidth,
    \item We find from the distribution of error introduced by lossy compression near matches to Laplacian distributions, which demonstrates a great potential for lossy compression to introduce differential privacy, an essential security technique for FL.
\end{itemize}

The organization of our paper is as follows. In Section \ref{sec:background}, we describe lossy compression's impact on communication overhead to motivate the study. In Section \ref{sec:related}, we discuss similar work in the field of FL communication cost reduction. Section \ref{sec:problem} describes our approach through mathematical formulation. Section \ref{sec:fedsz} describes our design strategies and methods for optimizing FL communications. We present our results in Section \ref{sec:evaluation}. We finally conclude the paper with a discussion and future directions in Section \ref{sec:con}.

\section{Background and Motivation}
\label{sec:background}
%FL is an ML framework designed to leverage distributed edge devices across a wide-area network (WAN) while maintaining data privacy~\cite{scale3}. Unlike traditional models, FL significantly reduces data volume and privacy concerns. In FL, edge clients perform local computations, lessening data transferred to the server, which cuts bandwidth consumption, latency, and storage needs~\cite{federated-applications}. Additionally, it enhances privacy since raw data remains on local devices, reducing the risk of data leaks~\cite{IIADMM}. However, FL framework still faces challenges, particularly data transfer bottlenecks, necessitating the exploration of data reduction methods like lossy compression to alleviate long I/O times for clients~\cite{challenges, scale3}. 
In this section we set the stage for how lossy compression can improve I/O efficiency in FL workflows and motivate our proposed \textsc{FedSZ} approach.

\subsection{Error-bounded Lossy Compression}\label{subsec:compression}
Lossy compression techniques explicitly designed for floating-point data like \texttt{SZ2}~\cite{sz2}, \texttt{SZ3}~\cite{sz3-1,sz3-2}, \texttt{SZx}~\cite{szx}, and \texttt{ZFP}~\cite{zfp} enable significant reductions in storage and transmission costs while preserving data accuracy needed for analysis. Below, we summarize four state-of-the-art EBLCs with fundamentally different designs and implementations. The four compressors are also representative of three classic lossy compression models: prediction-based, bit-wise encoding, and transform-based, respectively.

\begin{itemize}
    \item \textbf{\texttt{SZ2}}~\cite{sz2} operates on a prediction-based model. It processes datasets in small multi-dimensional blocks, applying hybrid prediction using Lorenzo and linear regression. Post-prediction, \texttt{SZ2} quantizes prediction errors and compresses the resultant integers using Huffman encoding and Zstd, ensuring efficient data reduction.
    \item \textbf{\texttt{SZ3}}~\cite{sz3-1,sz3-2} enhances the prediction model of \texttt{SZ2} with multi-dimensional dynamic spline interpolation, followed by quantization, Huffman encoding, and Zstd. This approach, not requiring storage of linear regression coefficients, offers improved compression ratios, especially beneficial for higher error bounds.
    \item \textbf{\texttt{SZx}}~\cite{szx}, designed for speed, adopts a bit-wise-operation-based encoding model. It segments data into consecutive blocks, determining if each can be represented as a constant block within a given error bound. Non-constant blocks undergo bit-wise truncation. This method's simplicity ensures rapid compression and decompression.
    \item \textbf{\texttt{ZFP}}~\cite{zfp}, diverging in approach, employs a transform-based model. It applies a custom orthogonal transform to data blocks, encoding the transformed coefficients with specialized bitplane encoders. \texttt{ZFP}'s method offers high compression ratios and speeds, benefiting from its optimized transform and encoding strategies.
\end{itemize}

\subsection{Compression on Networked Systems}
The efficiency of data compression in FL, particularly in the context of \textsc{FedSZ}, hinges on balancing the computational cost of compression and decompression against the time saved in data transmission. To formalize this, we define several key variables: \(t_C\) and \(t_D\) represent the runtimes for compression and decompression, respectively; \(S\) and \(S'\) are the original and compressed data sizes; and \(B_N\) signifies the network bandwidth. With these variables, we define the inequality in Eqn. \ref{eqn:comm-ineq},  which describes the situation where there is a runtime benefit from lossy compression: the total time spent on reduction (\(t_C\)), decompression (\(t_D\)), and transmitting the compressed data (\(S'/B_N\)) should be less than the time to send the original, uncompressed data (\(S/B_N\)).
\begin{equation}\label{eqn:comm-ineq}
    0 < t_C + t_D + \frac{S'}{B_N} < \frac{S}{B_N},
\end{equation}  
This inequality serves as a decision-making criterion, dictating when the benefits of reduced data transmission time outweigh the computational costs of compression. Its relevance to \textsc{FedSZ} lies in its ability to guide the algorithm toward `worthwhile' reduction, optimizing the trade-offs in scenarios with limited bandwidth and computational resources.

%\subsection{Federated Learning Data Management}
%FL involves collaboratively training a model across decentralized edge devices while keeping data localized. This preserves privacy but imposes statistical and systems-related challenges compared to centralized data settings~\cite{fedlit}. Statistically, FL must address non-IID (independent and identically distributed) data distributions across devices. This can hamper model convergence and accuracy compared to centralized datasets~\cite{fedlit}. Devices also have unbalanced and sparse data, which can skew training. A core constraint is the massive volume of data generated at the network edge. Sensors, multimedia, logs, and other sources produce enormous and rapidly growing data on resource-limited devices. Transmitting all this raw data to central servers is infeasible due to device capabilities, latency, and privacy risks. Thus, data must be locally stored and used for model training. However, edge devices have limited storage, bounded by megabytes to gigabytes. This severely restricts the volume of accumulated data for local learning. Smaller local datasets degrade model accuracy and training convergence compared to centralized settings.

These compression and communication challenges motivate solutions to efficiently manage and transfer data in FL without sacrificing model performance. Our proposed compression approach called \textsc{FedSZ}, described in later sections, directly addresses these needs to improve communication efficiency for the resource-constrained federated environment.

\section{Related Works}\label{sec:related}
In this section, we discuss three aspects of related work to identify the gap \textsc{FedSZ} aims to address while foreshadowing the advantages of our methodology.

\subsection{Impact of Lossy Compression on Data Transfer}

EBLC has been widely explored to improve data transfer performance significantly in many scenarios. For example, Liang et al.~\cite{data-dumping} found that various supercomputers may possess different I/O bandwidths in practice, and that selecting appropriate lossy compressors can be critical for maximizing the overall I/O performance. Specifically, Liang et al. characterized the I/O bandwidth on multiple supercomputers and developed an adaptive lossy compression framework combining \texttt{SZ} and \texttt{ZFP} to maximize the I/O performance. Liu et al.~\cite{Liu-ICDCS2023} developed \textit{Ocelet}, which can leverage parallel EBLC to accelerate the data transfer performance between Globus endpoints on a wide area network (WAN). Over 90\% of the transfer time can be reduced by Ocelet. Compared with the above existing use cases, using EBLC in FL faces more challenges such as lower network bandwidth between clients and servers and the parameter data being much less compressible than scientific datasets. Therefore, there is a gap in the known literature concerning how effective EBLC is in both edge case scenarios and FL scenarios.

\subsection{Compression of Machine Learning Models}
Several techniques have been proposed for compressing deep neural networks (DNNs), including lossless and lossy methods. Lossless compression approaches like lossless expressiveness optimization~\cite{lossless-comp} use linear programming and rectified linear units to encode network weights; however, this method can have low compression ratios and high compression time overheads. Lossy DNN compression techniques, such as DeepSZ~\cite{deepsz} and Delta-DNN~\cite{delta-dnn}, first apply pruning to sparsify networks and then quantization to shrink weight parameters. These lossy methods can often degrade model accuracy and/or require costly retraining to recover performance. In FL this is not tenable, as communication rounds can be infrequent and expensive for battery-constrained and distributed devices. Universal randomized lattice quantization~\cite{lossy-quant} has been proposed to compress DNNs by quantizing weight vectors, offering high compression ratios regardless of the distribution of parameters, with the limitation of low-granularity quantization. Therefore, it is necessary to explore EBLC's effects on model accuracy in FL settings to reduce the communication time overhead with high compression ratios while maintaining uncompressed accuracy.

\subsection{FL Communication Cost Reduction}
%Various approaches have emerged in the quest to mitigate communication costs in FL, each with its unique set of drawbacks. Kang and Ahn~\cite{kang} offer a method involving partial model structures sent to clients. However, this technique highly depends on specific neural network architectures and client resources, limiting its generalizability. Yang et al.~\cite{yang} propose RingFed, a decentralized aggregation method that, while practical, adds complexity and necessitates reliable local network conditions, potentially restricting its applicability in diverse settings. Wu et al.~\cite{wu} focuses on reducing the indexing overhead in sparse model updates with SmartIdx. This solution performs well for inherently sparse models but may not be universally applicable. These approaches, though innovative, often excel in specialized circumstances—be it model architecture, network topology, or update sparsity. Our work aims to offer a more universally applicable solution by integrating lossy compression to balance communication efficiency and model performance. 

In FL, various compression methods have been explored to enhance communication efficiency. Gradient sparsification, as initially proposed by Strom (2015), involves transmitting only significant gradients, with recent advancements favoring dynamic thresholds for broader applicability \cite{strom2015scalable, dryden2016communication, chen2018adacomp}. The Top-K method, a popular technique, selects the largest gradients for transmission, ensuring efficient parameter updates \cite{aji2017sparse, lin2017deep, sattler2019sparse, renggli2019sparcml}. Gradient quantization, exemplified by one-bit SGD and signSGD, reduces data transmission size but often results in biased estimates \cite{seide20141, bernstein2018signsgd}. Recent methods like TerGrad and QSGD employ stochastic unbiased estimations to maintain accuracy \cite{wen2017terngrad, alistarh2017qsgd}. Low-rank approaches, leveraging the inherent properties of over-parameterized DNN models, focus on efficient gradient matrix decomposition \cite{martin2021implicit, li2018algorithmic, vogels2019powersgd, yu2018gradiveq, wang2018atomo}. Comparison to the aforementioned existing methods is difficult or not possible as they are not open-source. Therefore, any meaningful comparison would require reimplementation, which is outside of the scope of this exploration.

While these methods offer potential benefits in FL, EBLC is able to reconstruct a dense network of weights with the original floating-point precision. \textsc{FedSZ} is a "last-step" in the communication pipeline, meaning that it is capable of further compressing sparsely trained or quantized model updates. Therefore, comparison to existing methods is not necessary, as any method can ostensibly can be used in concert with \textsc{FedSZ}. EBLC, through its reconstruction approach to compression and decompression, can help address the adjustments in training and sparsity that can introduce bias and inaccuracy in compressed models. 

\section{Problem Formulation}
\label{sec:problem}

In this section, we present the optimization framework that guides the design of \textsc{FedSZ}. The framework addresses two multi-objective optimization problems, aiming to improve both computational efficiency and model performance in an FL environment.

\subsection{Problem 1: Lossy Compressor Selection}

Our first problem focuses on selecting an optimal EBLC from a set, \( \mathcal{X}_{\text{lossy}} \). The objective is twofold: minimize the time overhead and maximize the compression ratio. These objectives are crucial as they directly affect the efficiency and speed of the \textsc{FedSZ} process. Each compressor \( x \in \mathcal{X}_{\text{lossy}} \) is associated with error bound \( \epsilon \), a critical parameter that impacts both the compression ratio \( R(x, \epsilon) \) and the runtime \( T(x, \epsilon) \).

% Justification for constraints
The constraints are specifically chosen to ensure practicality and efficiency. The feasible region for \( R \) is \( [1, S] \), with \( S \) being the original number of elements and 1 being no compression. This range ensures that the compression remains advantageous for a user. For the time \( T \), the feasible region is \( \left(0,\frac{S}{B_N}\right) \), where \( B_N \) is the network throughput, thus ensuring that the time overhead doesn't exceed network capabilities.

\begin{equation}\label{eqn:first-formulation}
\begin{aligned}
x^* = \arg \left\{ \max_{x \in \mathcal{X}_{\text{lossy}},\epsilon>0} R(x, \epsilon), \min_{x \in \mathcal{X}_{\text{lossy}},\epsilon>0} T(x, \epsilon) \right\} \\
\text{subject to} \quad
% \epsilon > 0, \\
0 < T(x, \epsilon) < \frac{S}{B_N}, \\
1 \leq R(x, \epsilon) \leq S.
\end{aligned}
\end{equation}

\subsection{Problem 2: Compression in Federated Learning}

Building upon Problem 1, our second research problem focuses on effectively integrating our optimal compressor \( {x}^* \) into the \textsc{FedSZ} algorithm. The choice of \( {x}^* \) directly influences the computational overhead, communication reduction, and accuracy impacts in our distributed FL setting, making it a cornerstone for this problem.

Here, we suppose there are \( n \) clients, and each client-$i$ has an associated communication costs \( P_i \) and network bandwidth \( B_{N,i} \). The error bound variable \( \epsilon \) is our critical factor, affecting both the communication cost \( P_i(\epsilon) \) and the inference accuracy \( I(\epsilon) \) at the server during aggregation and validation. Parameter \( I'\in[0,1] \) serves as the baseline accuracy when no compression is applied, and the goal is to find the optimal $\epsilon^*$ that minimizes the discrepancy between \( I' \) and \( I(\epsilon) \).

% Constraints and their justification
The constraints are set to ensure that the solutions are both feasible and effective. The error bound \( \epsilon \) is constrained to be greater than zero to avoid trivial solutions. The communication cost \( P_i(\epsilon) \) is bounded by the network bandwidth, and the accuracy \( I(\epsilon) \) is naturally bounded between 0 and 1 to represent valid probability measures.

\begin{equation}\label{eqn:second-formulation}
\begin{aligned}
\epsilon^* = \arg \left\{ \min_{\epsilon > 0} \sum_{i = 1}^n P_i(\epsilon), \ \min_{\epsilon > 0} \left|I' - I(\epsilon) \right| \right\} \\
\text{subject to} \quad
0 \leq I(\epsilon) \leq 1, \\
0 \leq P_i(\epsilon) \leq \frac{S_i}{B_{N,i}}.
\end{aligned}
\end{equation}

Formulations \eqref{eqn:first-formulation} and \eqref{eqn:second-formulation} provide the mathematical foundation for the \textsc{FedSZ} algorithm. These equations tackle the computational and communication complexities in FL, thereby laying the groundwork for the subsequent development and analysis of the \textsc{FedSZ} algorithm.

\section{FedSZ: A Federated Lossy Compression Scheme}
\label{sec:fedsz}
\begin{figure*}[!htb]
  \centering  
\subfigure[{Compression Pipeline at Client}]
{
\label{fig:sub1}
\raisebox{-1cm}{\includegraphics[scale=0.54]{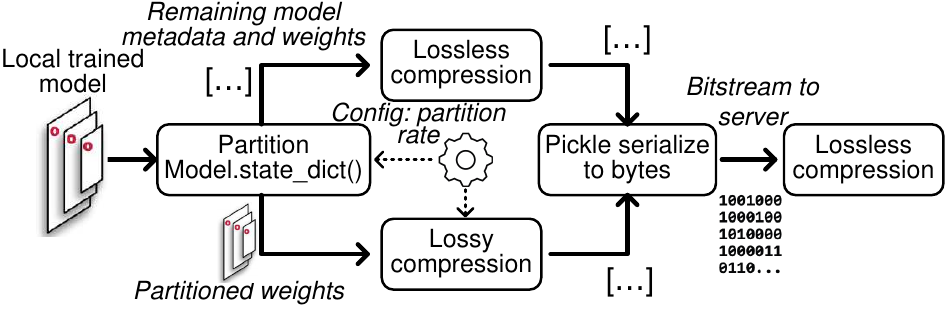}}
}  
\subfigure[{Decompression Pipeline at Server}]
{
\label{fig:sub2}
\raisebox{-1cm}{\includegraphics[scale=0.54]{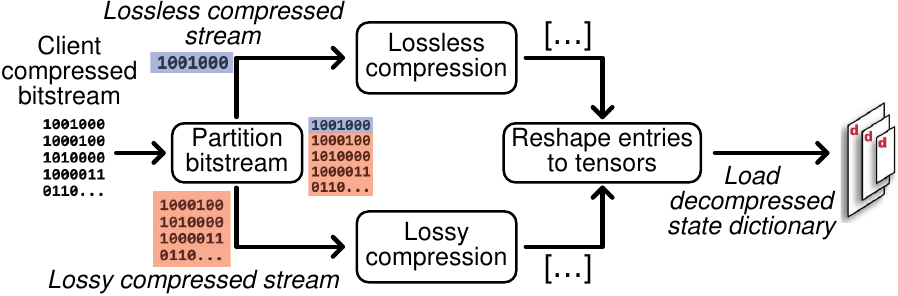}}
}  

  \caption{\textsc{FedSZ} Design for both Compressing and Decompressing Local Model Updates}
  \label{fig:combined}
  \vspace{-3mm}
\end{figure*}

Our main contribution in this paper is \textsc{FedSZ}: a generally applicable compression scheme for FL client-server communications. Our technique is summarized as follows: (i) partitioning a client update (represented as a PyTorch model state dictionary) into lossy and lossless components, (ii) lossy and lossless compression of the partitions, and (iii) communication of the bitstream for decompression at the receiving host. This method is illustrated in Figure \ref{fig:combined}, which involves the compression stage (as shown in Figure \ref{fig:sub1}) and decompression stage (as shown in Figure \ref{fig:sub2}). The rest of this section details the different considerations of our design and our strategies to solve our outlined research problems from Section~\ref{sec:problem}.

\subsection{Characterizing Model State Data}\label{subsec:bad-data}

\begin{figure}[ht] 
\vspace{-0.5em}
\centering
\captionsetup{justification=centering}
\hspace{-10mm}
\subfigure[{Snippets of FL Model Parameters}]
{
\raisebox{-1cm}{\includegraphics[scale=0.36]{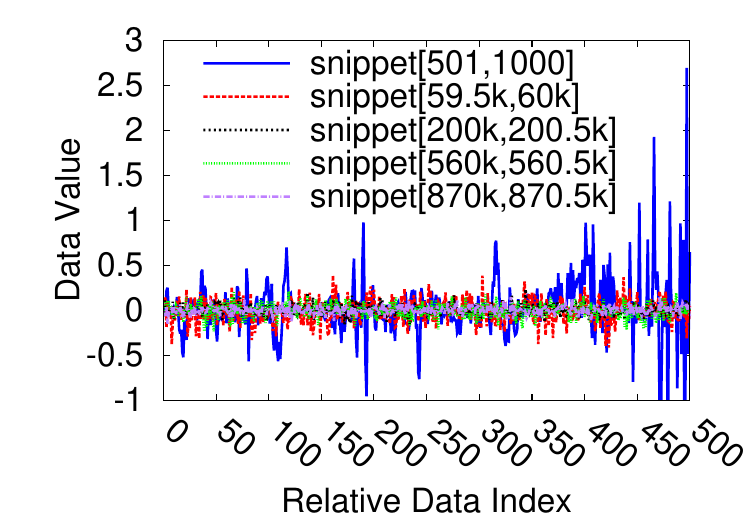}}
}
\hspace{-8mm}
\subfigure[{Snippets of Miranda Data}]
{
\raisebox{-1cm}{\includegraphics[scale=0.36]{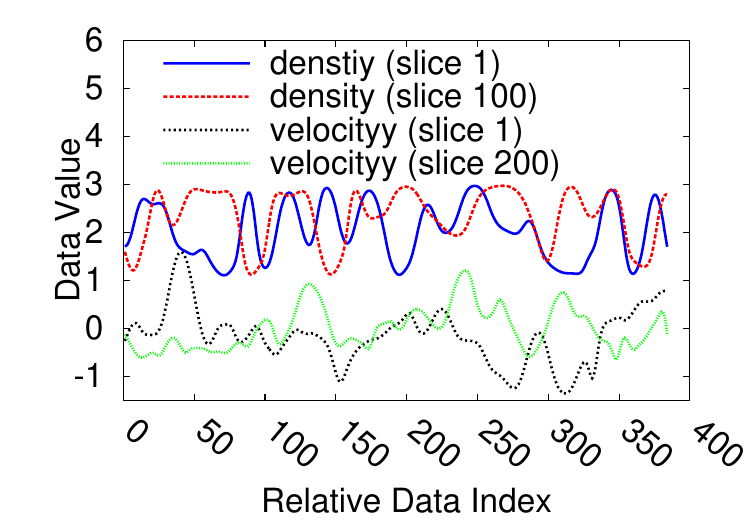}}
}
\hspace{-10mm}

%\vspace{-4mm}
\hspace{-6mm}
\subfigure[{Density visualization}]
{
\raisebox{-1cm}{\includegraphics[scale=0.18]{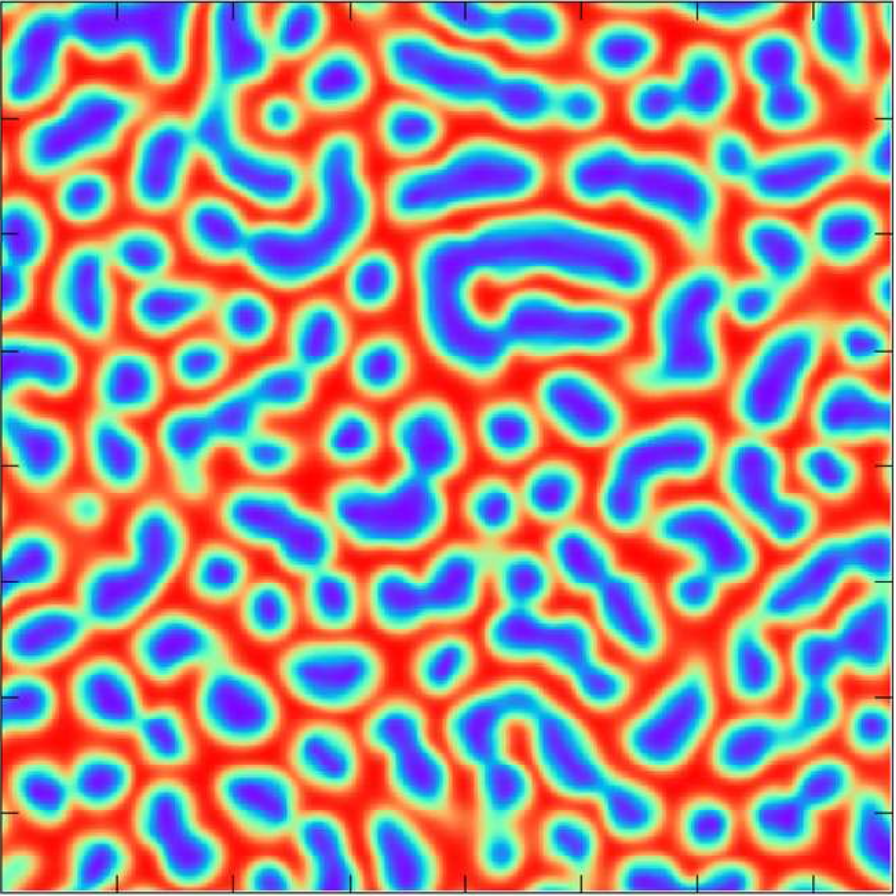}}
}
\hspace{5mm}
\subfigure[{Velocity visualization}]
{
\raisebox{-1cm}{\includegraphics[scale=0.18]{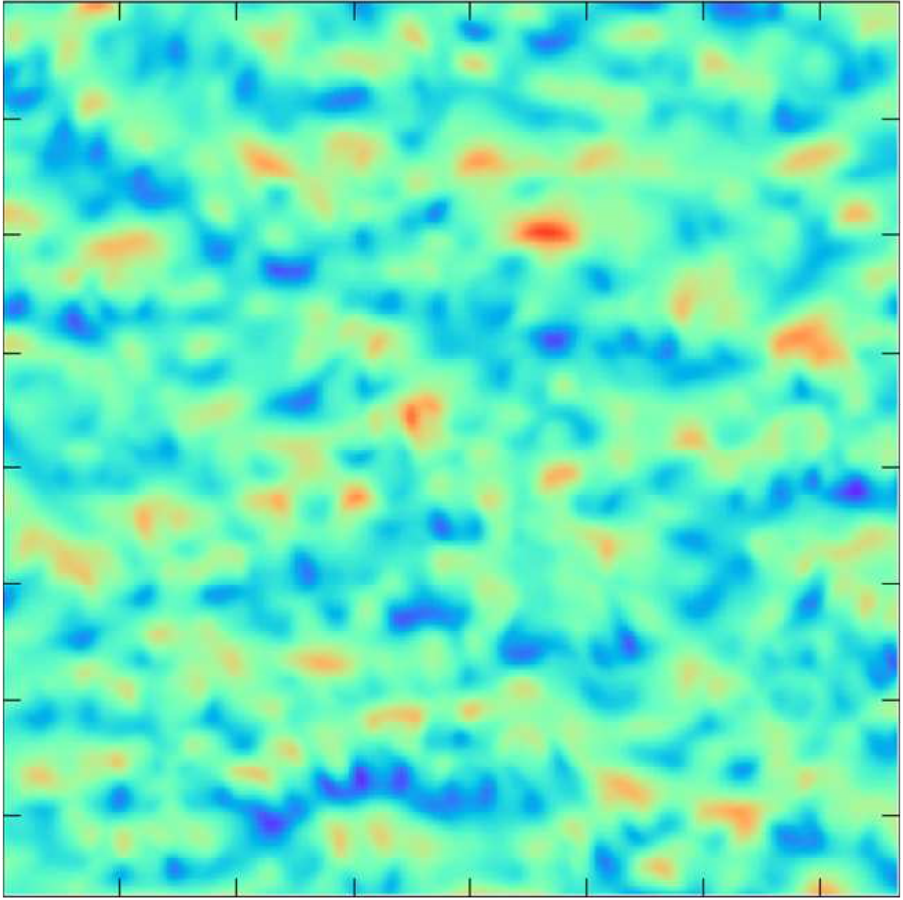}}%
}
\hspace{-10mm}
\vspace{-3mm}

\caption{Comparing FL Model Parameters vs. Scientific Simulation Data (MIRANDA~\cite{miranda, sdrbench})}
\label{fig:comparison}
\vspace{-4mm}
\end{figure}

First, we characterize the shape and distribution of FL model weights in this subsection, something that is critical to understanding the challenges of using EBLC to compress model data. We compare the value variation of FL model parameters versus the classic scientific simulation data in Figure \ref{fig:comparison}, where the relative data index means the 1D data index in the specific snippet or slice. The figure illustrates that the FL model parameters are very spiky, while the classic simulation datasets exhibit much higher smoothness. This is because the scientific simulation data are often used to convey specific physical or chemical phenomena, as demonstrated in Figure \ref{fig:comparison} (c) and (d).
As such, a serious question arises: Can the EBLCs still work effectively on FL model parameters? Moreover, which compressor is the best choice? We answer them in our study.

\subsection{Overview of \textsc{FedSZ} Algorithm}\label{sec:fedsz-alg}

We design \textsc{FedSZ} based on observations from careful analysis and experiments. Specifically, we learn that our compression scheme, summarized in Figure \ref{fig:combined}, should include three strategies: (i) partitioning the client update into lossy and lossless components, (ii) using \texttt{SZ2} to lossy compress and \texttt{blosc-lz} to lossless compress, and (iii) dumping a bitstream to communicate the client update to the server. 
%The algorithmic representation of the compression scheme is shown below. 
We describe the design details in the Algorithm \ref{alg:FedSZ}.
\begin{algorithm}[ht]
 \caption{\textsc{FedSZ} Compression Scheme}
 \label{alg:FedSZ}
 \footnotesize
 %\fontsize{10pt}{11pt}\selectfont
\begin{algorithmic}[1]
\REQUIRE model: \textit{torch.Module}, threshold: \textit{int}
\ENSURE \textit{compressed\_model: bytes}
\STATE \textbf{Initialize}: \textit{compressed\_params} = \{\}
\FOR{\textbf{each} (name, param) \textbf{in} model.state\_dict()}
    \STATE flat\_param $\leftarrow$ param.flatten()
    \IF{``weight" \textbf{in} name \textbf{and} flat\_param.size $>$ threshold}
        \STATE compressed\_param $\leftarrow$ lossy\_compress(flat\_param)
    \ELSE
        \STATE compressed\_param $\leftarrow$ lossless\_compress(flat\_param)
    \ENDIF
    \STATE \textit{compressed\_params[name]} $\leftarrow$ compressed\_param
\ENDFOR
\RETURN \textit{compressed\_params}.to\_bytes()
\end{algorithmic}
\end{algorithm}

\subsection{Partitioning the State Dictionary}

Our \textsc{FedSZ} algorithm employs a partitioning strategy dividing the state dictionary into the components that we can compress without sacrificing accuracy and the parts that require lossless compression to maintain the integrity of the model state (as shown in Line 4 of Algorithm \ref{alg:FedSZ}). 
The complexities in efficiently compressing DNNs necessitate a method for partitioning the model's state dictionary into lossy and lossless compressible segments (see line 4-8). In a DNN, maintaining a model for training and evaluation involves various layers and values, including parameter tensors, running means, and bottlenecks. The \verb|state_dict()| captures the complete state of the model, both trainable and non-trainable. However, lossy compression of both parameters/weights and metadata risks significant loss of important values and extreme degradation of model accuracy, which has we have experimentally verified. Our observation is also consistent with DeepSZ~\cite{deepsz}, which aims to save storage space for ML models using lossy compression. As such, we adopt an selective compression strategy to partition models into the dense, compressible weights and non-compressible metadata. 

\subsection{Exploring the Most Effective Lossy Compression Method}\label{subsec:lossy-comp-methods}
 In our exploration of determining which lossy compressor we should use, we compare \texttt{SZ2} (v1.12.5), \texttt{SZ3} (v3.1.7), \texttt{SZx} (v1.0.0), and \texttt{ZFP} (v1.0.0), all EBLCs with different characteristics as described in Section \ref{subsec:compression}. To decide which EBLC to use, we carefully investigate (i) the distribution of the data to compress, (ii) the impact on inference accuracy, and (iii) the performance of the compressors. All runtime and throughput data are computed on a Raspberry Pi 5 with 8GB of RAM. All training and inference accuracy data are computed on Argonne's Swing cluster, where each node has 8$\times$NVIDIA A100 40GB GPUs and 2$\times$AMD EPYC 7742 64-Core Processors with 128 cores.
\subsubsection{Relative Error Bounding}
\begin{figure}[!htb]
    \centering
    \includegraphics[width=\columnwidth]{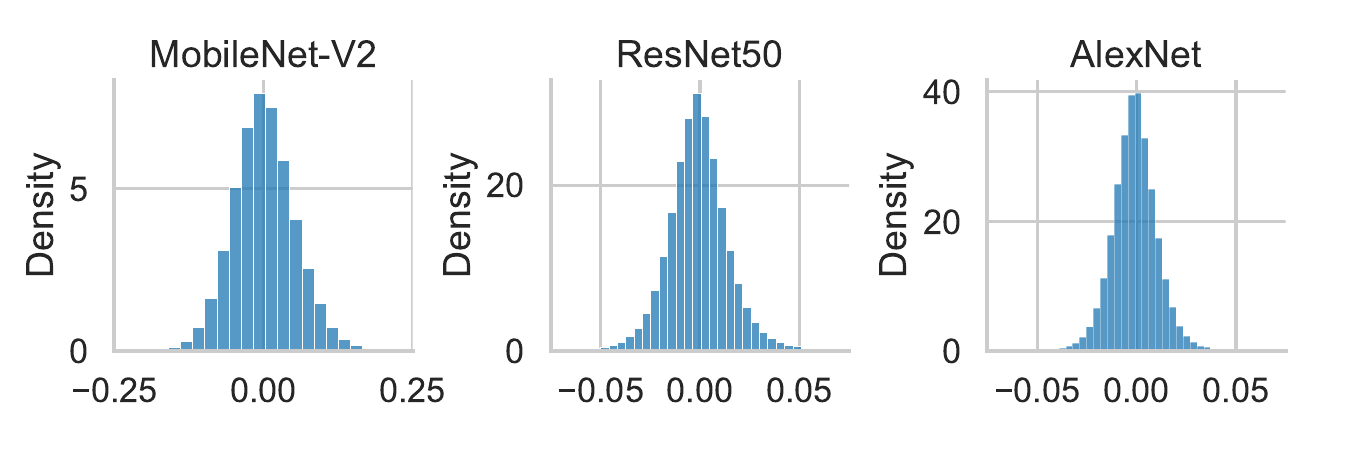}
    \caption{Distribution of Pretrained Weights for Various Models}
    \label{fig:distribution}
    \vspace{-5mm}
\end{figure}

Selecting an appropriate error bounding mode is a nuanced task that directly correlates with the statistical characteristics of the data to be compressed~\cite{liu2023srnsz}. Our study focuses on the weight distributions of three distinct models: AlexNet, MobileNetV2, and ResNet50, as visualized in Figure \ref{fig:distribution}. While each model's weight distribution is between -1 and 1, they exhibit different dynamic ranges and clustering behaviors around zero. 

Given these variations in weight distributions, a fixed error bound could either be overly conservative for some data segments or too liberal for others, thereby not exploiting the compression potential to its fullest. This is where the utility of relative error bounds becomes apparent. A \textit{relative error bound} adapts to the local properties of the data by being a multiple of the dynamic range, ensuring that the error introduced during compression is proportional to the data's actual value. Consequently, we opt for relative error bounding modes for \texttt{SZ2}, \texttt{SZ3}, and \texttt{SZx}. For \texttt{ZFP} there is not a relative error mode, so we select the closest analogous option, which is fixed precision mode, where the number of uncompressed bits is fixed. This choice allows us to harness the benefits of adaptive error control, accommodating the dynamic ranges observed across different layers in a model.

\subsubsection{Inference Accuracy Convergence Comparison} 
\begin{figure}[ht]
    \centering
    \includegraphics[width=\columnwidth]{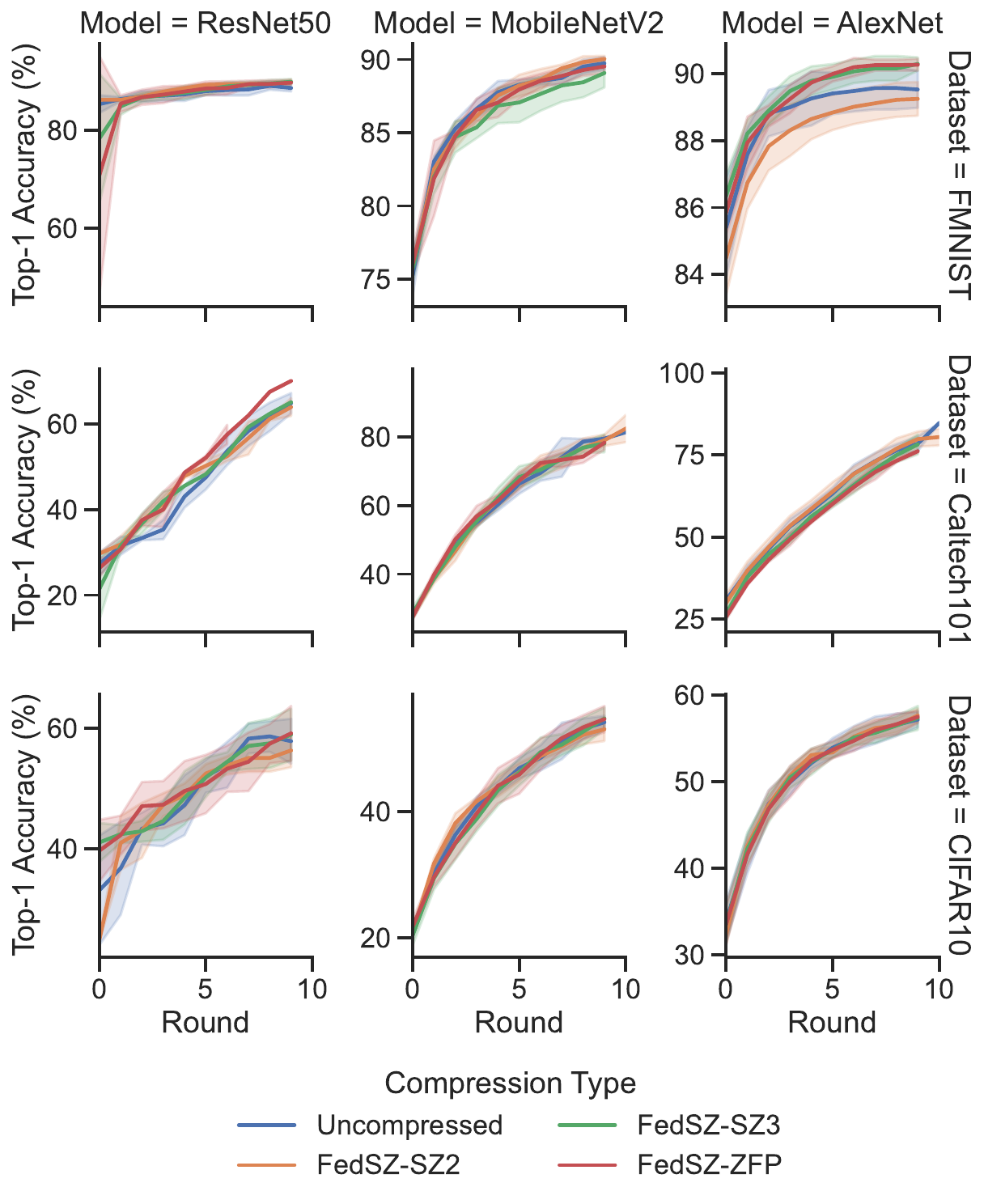}
    \caption{Accuracy Convergence Comparison for EBLCs}
    \label{fig:lossy-accuracy}
    \vspace{-3mm}
\end{figure}

The measure of the efficacy of a lossy compression algorithm in an FL setting such as \textsc{FedSZ} is not just compression performance but also its impact on the model's inference accuracy and convergence behavior. Achieving a high compression ratio is of little benefit if it comes at the cost of a model that fails to converge to a satisfactory level of accuracy. Therefore, in addition to comparing raw compression metrics, we also closely examine how each compressor influences the model's accuracy throughout communciation rounds. We run an FL simulation on a cluster of training AlexNet for the CIFAR-10 task for ten rounds with one epoch per round while compressing the updates with each candidate compressor. We note that results for other models and datasets are similar as it is essentially compression of "spiky" 1D floating-point data, therefore other results are not included. The observed trends in inference accuracy for these compressors are depicted in Figure \ref{fig:lossy-accuracy}.

\begin{table*}[htb]
\centering
\caption{EBLC Comparison Across Different Models for CIFAR-10--Throughput refers to an EBLC's data processing rate}
\resizebox{\linewidth}{!}{
\begin{tabular}{llcccccccccccc}
\toprule
        &  &  \multicolumn{3}{c}{Runtime (s)} & \multicolumn{3}{c}{Throughput (MB/s)} & \multicolumn{3}{c}{Compression Ratio} & \multicolumn{3}{c}{Top-1 Accuracy (\%)}  \\
\cmidrule(lr){3-5} \cmidrule(lr){6-8} \cmidrule(lr){9-11} \cmidrule(lr){12-14}
             &            &\multicolumn{12}{c}{Relative Error Bound} \\
             Model &      Compressor      & $10^{-2}$ & $10^{-3}$ & $10^{-4}$   & $10^{-2}$ & $10^{-3}$ & $10^{-4}$ & $10^{-2}$ & $10^{-3}$ & $10^{-4}$ & $10^{-2}$ & $10^{-3}$ & $10^{-4}$  \\
\midrule
\multirow{4}{*}{AlexNet}  
             & \texttt{SZ2}        & $3.22$   & $4.93$   & $6.64$   & $70.75$  & $46.26$  & $34.34$  & $\mathbf{11.26}$ &$\mathbf{5.228}$ & $\mathbf{3.375}$ & $\mathbf{57.90}$ & $\mathbf{57.39}$ & $57.06$ \\
             & \texttt{SZ3}        & $7.22$   & $8.80$   & $10.69$  & $31.58$  & $25.94$  & $21.34$ & $9.827$ & $5.169$ & $3.345$ & $57.28$ & $57.18$ & $\mathbf{58.50}$  \\
             & \texttt{SZx}        & $\mathbf{0.444}$ & $\mathbf{0.439}$ &$\mathbf{9.445}$ & $\mathbf{3514.92}$ & $\mathbf{3554.84}$ & $\mathbf{3507.02}$ &$4.804$ & $4.801$ & $4.802$ & $10.00$ & $10.00$ & $10.00$ \\
             & \texttt{ZFP}        & $1.89$   & $2.11$   & $2.36$   & $120.66$ & $108.17$ & $96.51$ & $4.166$ & $2.942$ & $2.210$ & $57.30$ & $56.80$ & $57.19$  \\
             \midrule
\multirow{4}{*}{MobileNet-V2}
             & \texttt{SZ2}        & $0.365$  & $0.482$  & $0.692$  & $25.61$  & $19.38$  & $13.51$ & $\mathbf{5.409}$ &$3.235$ & $1.923$ & $\mathbf{55.19}$ & $55.19$ & $\mathbf{55.19}$\\
             & \texttt{SZ3}        & $0.574$  & $0.724$  & $1.082$  & $16.30$  & $12.92$  & $8.64$ & $5.250$ & $3.125$ & $1.779$ & $54.94$ & $56.14$ & $53.86$ \\
             & \texttt{SZx}        & $\mathbf{0.034}$ & $\mathbf{0.031}$ & $\mathbf{0.029}$ & $\mathbf{104.089}$ & $\mathbf{114.162}$ & $\mathbf{122.036}$&  $4.799$ & $\mathbf{4.765}$ & $\mathbf{4.783}$ & $10.00$ & $10.00$ & $10.00$  \\
             & \texttt{ZFP}        & $0.127$  & $0.138$  & $0.151$  & $73.73$  & $67.57$  & $61.70$ & $3.027$ & $2.333$ & $1.897$ & $54.600$ & $\mathbf{57.26}$ & $54.90$ \\
            \midrule
\multirow{4}{*}{ResNet50}    
             & \texttt{SZ2}        & $1.235$  & $1.930$  & $2.870$  & $76.80$  & $49.14$  & $33.05$ & $\mathbf{7.025}$ & $4.041$ & $2.737$ & $58.66$ & $59.15$ & $58.66$  \\
             & \texttt{SZ3}        & $2.723$  & $3.459$  & $4.586$  & $34.82$  & $27.42$  & $20.68$ & $6.768$ & $3.942$ & $2.662$ & $\mathbf{64.14}$ & $61.35$ & $63.51$ \\
             & \texttt{SZx}        & $\mathbf{0.186}$ & $\mathbf{0.186}$ &$\mathbf{0.185}$ & $\mathbf{3516.02}$ & $\mathbf{3516.02}$ & $\mathbf{3535.27}$ &$4.806$ & $\mathbf{4.806}$ & $\mathbf{4.806}$ & $10.00$ & $10.00$ & $10.00$ \\
             & \texttt{ZFP}        & $0.747$  & $0.843$  & $0.962$  & $126.95$ & $112.53$ & $98.58$ & $3.449$ & $2.562$ & $2.035$ & $62.07$ &$\mathbf{64.70}$ & $\mathbf{64.06}$  \\
\bottomrule
\end{tabular}}
\label{table:multi_model_comparison}
\vspace{-5mm}
\end{table*}

As illustrated, the accuracy convergence for most compressors are closely aligned, indicating a minimal difference in compressor impact. The outlier compressor for accuracy retention is \texttt{SZx}, which fails to maintain any accuracy, likely due to its block mean storage. Our findings affirm that a nuanced balance between compression performance and accuracy impact is critical for the practical applicability of the chosen lossy compressor.

\subsubsection{Performance Characteristics}
We summarize our comparison of EBLCs for \textsc{FedSZ} in Table \ref{table:multi_model_comparison} when using a Raspberry Pi 5 for model compression. From these results, we find that \texttt{SZ2} is the optimal compressor. This table summarizes key metrics across multiple error bounds. The accuracy metrics has been evaluated on a cluster while training the AlexNet, MobileNet-V2, and ResNet50 models on the CIFAR-10 dataset over ten communication rounds, and the runtime and throughput data is from compressing the trained models on a Raspberry Pi 5. There was no impact on training time besides the incurred overhead of compression, as will be discussed in Section~\ref{sec:time-gains}.

Our primary focus is discerning which EBLCs achieve good compression ratios with a low runtime and then maintain accuracy similar to not using compression. In this regard, \texttt{SZ2} stands out. It outperforms \texttt{SZ3} and \texttt{ZFP} in compression ratios by $14.6\%$ and $170.3\%$, respectively for AlexNet. Notably, this comes at the relatively modest increase in runtime when compared to \texttt{ZFP}, only $3.1\%$ more per epoch. \texttt{SZ2} continues to demonstrate a strong balance between high compression ratios and maintaining high accuracy across different models. For MobileNet-V2, at an error bound of 1E-2, \texttt{SZ2} achieves a compression ratio of $5.409$ and a Top-1 Accuracy of $55.19\%$, outperforming \texttt{SZ3} and \texttt{ZFP} in both metrics. Similarly, for ResNet50, \texttt{SZ2} achieves a compression ratio of $7.025$ and maintains a Top-1 Accuracy of $58.66\%$, surpassing its counterparts. For reducing I/O overhead, a higher compression ratio can be more valuable than marginal gains in runtime~\cite{barr}, and \texttt{SZ2}'s modestly greater runtime is a justifiable tradeoff for its superior compression ratio. 

The justifications for these results are as follows. Since FL model parameters are flattened to 1D data arrays with an unpredictable distribution, (as demonstrated in Section \ref{fig:comparison}) after which we have to perform 1D lossy compression. \texttt{ZFP} is optimized particularly for multi-dimensional datasets but may perform poorly on 1D spiky datasets, as disclosed by prior literature~\cite{nbody,nbody-dingwen-17, MDZ}. Thus, it achieves relatively low compression ratios on FL model parameters. \texttt{SZx}~\cite{szx} uses relatively simple compression operations (such as marking constant blocks and using bit-wise operations) as it was designed primarily to have low compression/decompression runtimes, which, may cause poor compression ratios and reconstructed data quality. \texttt{SZ2} and \texttt{SZ3} should exhibit similar compression ratios because they both default to using a Lorenzo predictor and quantization when data exhibit significant variations, according to their hybrid compression design~\cite{sz2,sz3-1}. In comparison, \texttt{SZ3} features a more sophisticated predictor selection policy, leading to lower compression throughput. All in all, \texttt{SZ2} exhibits the best tradeoff in terms of runtime, compression ratio, and data reconstruction quality, which we verify through our result in Table \ref{tab:compression_ratios}. 

Our method for selecting \texttt{SZ2} as our preferred compressor aligns coherently with the design criteria and feasibility region outlined in Eqn. \eqref{eqn:first-formulation}. Here, we are guided by the dual objectives of achieving a low runtime and a high compression ratio, all within a predefined feasibility region. The empirical data solidifies \texttt{SZ2} as the most suitable lossy compressor for \textsc{FedSZ}, effectively balancing performance metrics with the practical constraints of an FL environment.

\subsection{Exploring Best Fit Lossless Compression for Metadata and Non-weight Parameters}
In the \textsc{FedSZ} algorithm, we use lossless compression to reduce the size of client updates' metadata and non-weight parameters ($\approx1\%$ of an update's storage size). Similar to the lossy case, we evaluate several state-of-the-art lossless compressors. The considered compressors include \texttt{blosc-lz} (v1.21.3)~\cite{blosc}, \texttt{zlib} (v1.2.13)~\cite{zlib}, \texttt{zstd} (v1.5.5)~\cite{zstd}, \texttt{gzip} (Python v3.11.4)~\cite{gzip}, and \texttt{xz} (v0.5.0)~\cite{xz}. To compare these options, we compress the metadata and non-weight parameters for AlexNet to evaluate the compressors' performances, including the results in Table \ref{table:lossless_comparison}. Since lossless compressors do not introduce noise into the client updates, we do not need to check if there are impacts on inference accuracy.
% Table for Lossless Compressors
\begin{table}[h]
\centering
\captionsetup{justification=centering}
\caption{Lossless Compressor Comparison for Compressing AlexNet Metadata on Raspberry Pi 5}
\begin{tabular}{lccc}
\toprule
Compressor & Runtime (s) & Throughput (MB/s) & Compression Ratio  \\
\midrule
\texttt{blosc-lz}   & $\mathbf{0.271}$     & \textbf{$\mathbf{674.5}$}             & 1.248                           \\
\texttt{gzip}       & $7.728$       & $28.16$             & $1.160$                             \\
\texttt{xz}         & $74.52$       & $4.00$              & $\mathbf{1.250}$                         \\
\texttt{zlib}       & $7.772$       & $28.37$             & $1.164$                       \\
\texttt{zstd}       & $0.529$       & $348.6$             & $1.169$                             \\
\bottomrule
\end{tabular}
\label{table:lossless_comparison}
\end{table}

From Table~\ref{table:lossless_comparison}, we notice that \texttt{blosc-lz} outperforms the other lossless compressor by achieving more than $2 \times$ lower runtime than \texttt{zstd} and a comparable compression ratio to \texttt{xz}, a very slow lossless compressor. Taking these two points, it is clear to see that \texttt{blosc-lz} is a fair choice for our metadata. Just like the data distribution mentioned in Section~\ref{subsec:bad-data}, it is important to note that the target data are formatted as floating-point, small-size 1D arrays of non-uniform data, which will lead to low compressibility. 

\begin{figure*}[!htb]
  \centering
\subfigure[{Caltech101}\hspace{-5mm}]
{
\label{fig:caltech-error}
\raisebox{-1cm}{\includegraphics[scale=0.24]{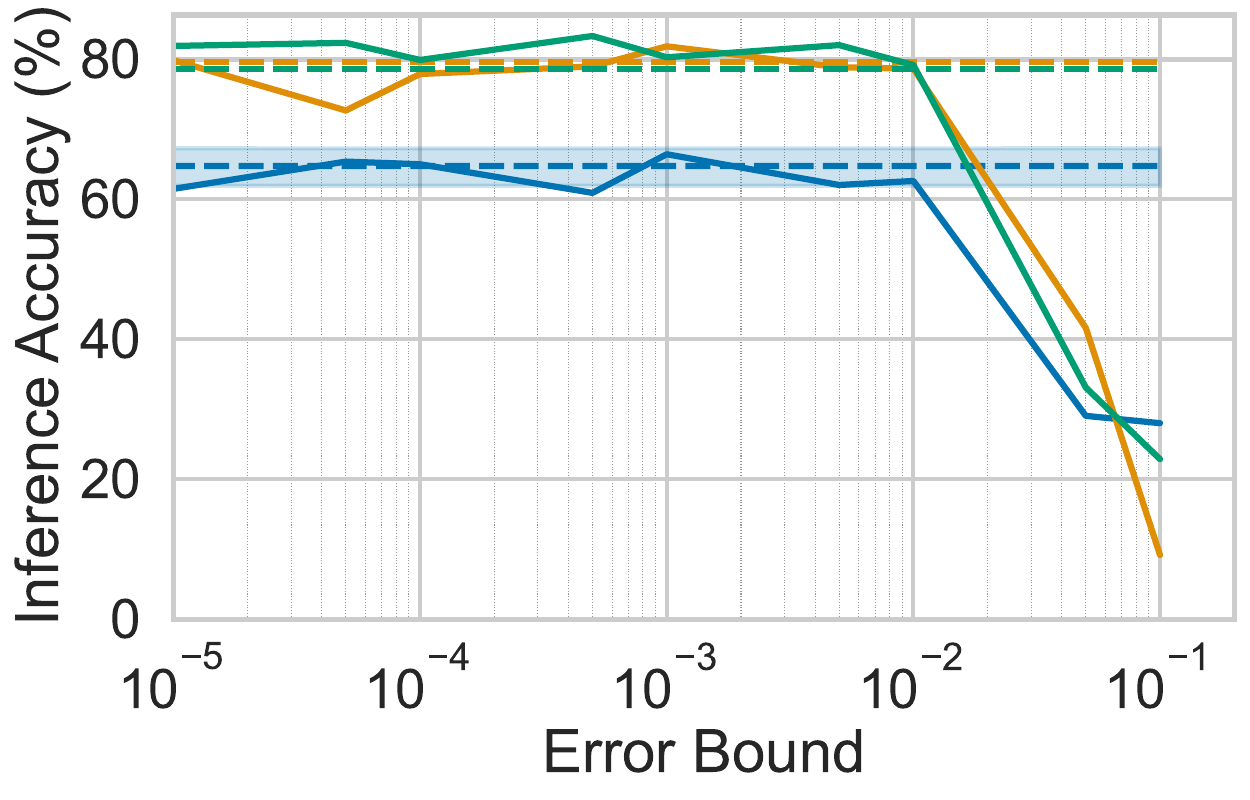}}
}
\subfigure[{CIFAR-10}\hspace{-5mm}]
{
\label{fig:cifar-error}
\raisebox{-1cm}{\includegraphics[scale=0.24]{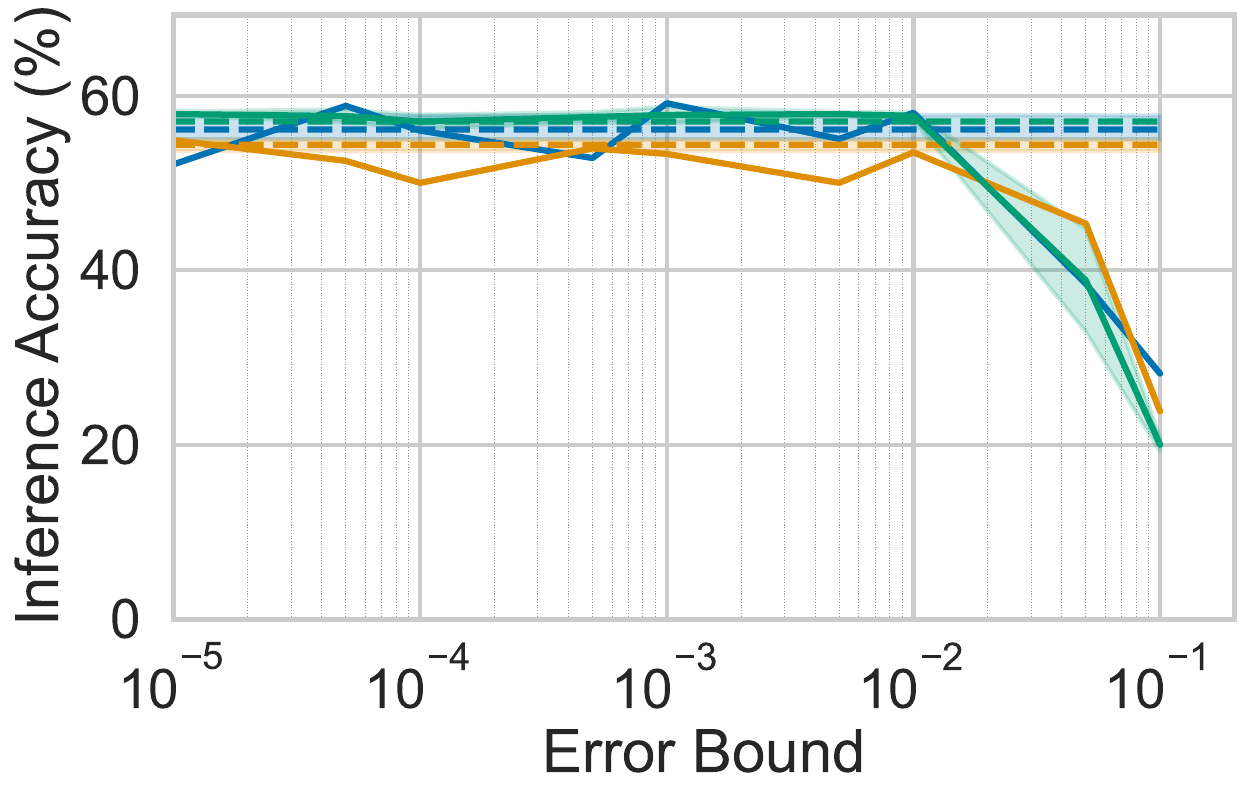}}
}
\subfigure[{Fashion-MNIST}\hspace{17mm}]
{
\label{fig:fmnist-error}
\raisebox{-1cm}{\includegraphics[scale=0.24]{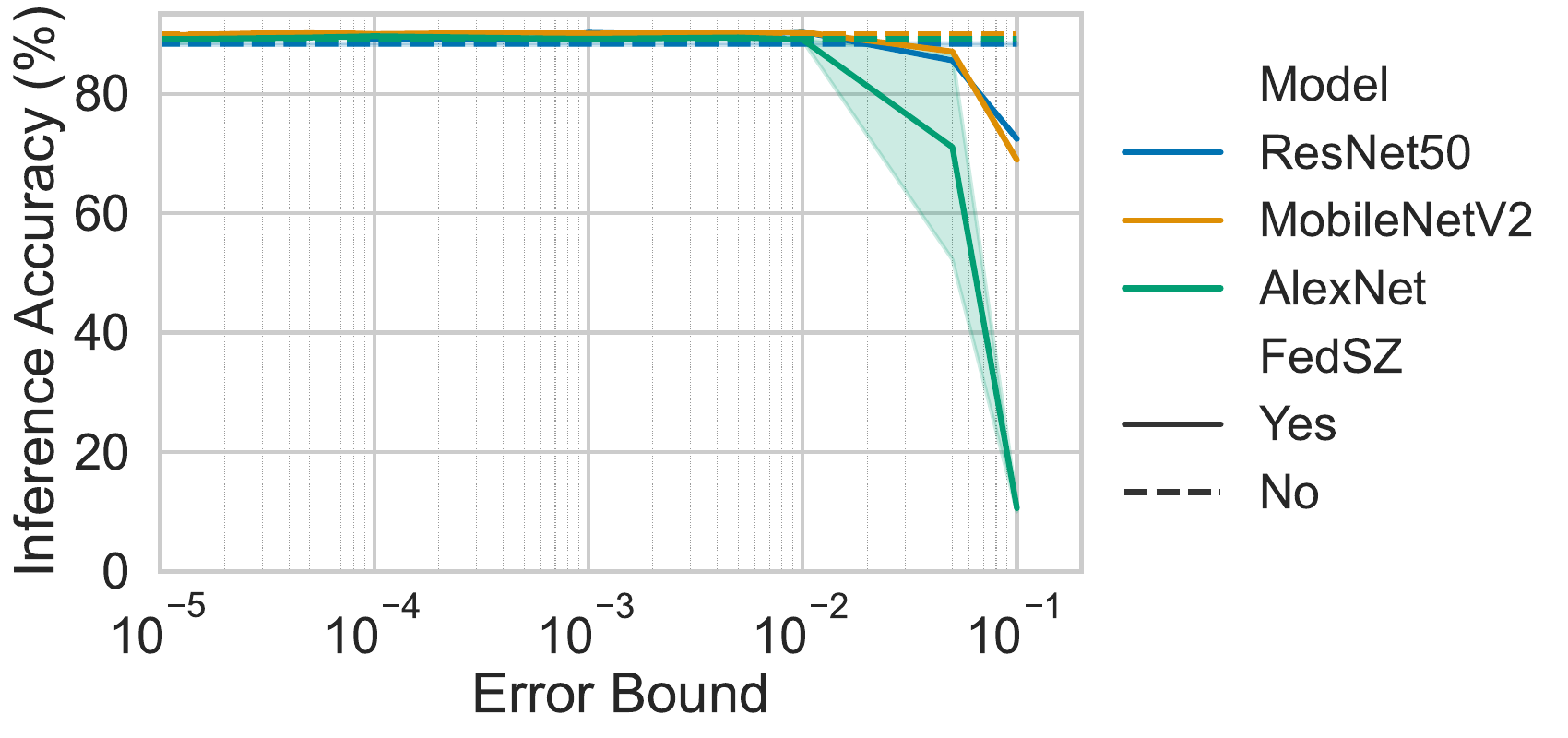}}
}
\vspace{-3mm}
  \caption{Inference Accuracy Across Diverse Models and Datasets while Varying \textsc{FedSZ} Relative Error Bound}
  \label{fig:error-overheads}
  \vspace{-4mm}
\end{figure*}

\section{Methodology for Evaluating \textsc{FedSZ}}
\label{sec:evaluation}
With the design of \textsc{FedSZ} documented, we now detail our evaluation methodology.
\subsection{Federated Learning Platform: APPFL with FedAvg}
We implement \textsc{FedSZ} in an open-source Python package, APPFL (v0.4.0)~\cite{appfl}, a library designed for the development and evaluation of privacy-preserving FL (PPFL) algorithms. APPFL offers modular APIs for implementing essential components such as learning algorithms, privacy mechanisms, communication protocols, models, and data handling. The package utilizes gRPC for cross-platform communication and employs Message Passing Interface (MPI) through mpi4py~\cite{mpi4py} for parallelism to enable scalable simulations. It is compatible with PyTorch, allowing for the integration of custom neural network models. APPFL has been successfully used to train models across various decentralized datasets, including biomedical images~\cite{hoang2023enabling} and electric grid~\cite{bose2023federated,bose2023privacy}.

In FL, the package provides the capability to train a global model by aggregating updates from client models operating on decentralized data. Various algorithms can be employed, each with different privacy, efficiency, and robustness trade-offs. For the scope of this study, we focus on Federated Averaging (FedAvg)~\cite{fedavg}, a well-established algorithm that performs local Stochastic Gradient Descent (SGD) on client devices and averages these local models to update a global one. The choice of FedAvg is deliberate; its simplicity, scalability, and robust performance make it compatible with compression techniques, thus serving our study's objectives effectively.

\subsection{Models and Datasets for Training}
The primary aim of \textsc{FedSZ} is to optimize client-server communication efficiency without sacrificing model accuracy or imposing a significant compression runtime overhead. To rigorously evaluate our framework, we chose DNNs that vary significantly in terms of parameter count, model size, percentage of data that we lossy compress, and computational complexity (FLOPs). The characteristics of these models are summarized in Table \ref{tab:models}.
\begin{table}[!htb]
    \centering
    \caption{DNNs for \textsc{FedSZ} Profiling: Mean Statistics}    
    \resizebox{\columnwidth}{!}{
    \begin{tabular}{ccccc}
    \toprule
        \textbf{Model} & \textbf{Parameters} & \textbf{Size} & \textbf{\% Lossy Data} & \textbf{FLOPs} \\
    \midrule
    MobileNet-V2~\cite{mobilenetv2} & $3.5\si{e}+06$ & $14\si{MB}$ &$96.94\%$ & $0.35\si{G}$ \\
    ResNet50~\cite{resnet18} & $4.5\si{e}+07$ & $180\si{MB}$ &  $99.47\%$ & $8\si{G}$ \\
    AlexNet~\cite{alexnet} & $6.0\si{e}+07$ & $230\si{MB}$ & $99.98\%$ & $0.75\si{G}$ \\
    \bottomrule
    \hspace{1mm}
    \end{tabular}}
    \label{tab:models}
 \vspace{-3mm}
\end{table}

Selecting a diverse set of models ensures that our results are not specific to any particular architecture, making our insights broadly applicable. For instance, MobileNet-V2, with its relatively fewer parameters and FLOPs, represents edge cases where the device capabilities might be limited. In contrast, ResNet50 offers a more complex architecture suitable for resource-rich environments, and AlexNet serves as a middle-ground model. 

Similarly, we chose three well-established image classification datasets to maintain comparable tasks across different models, as detailed in Table \ref{tab:datasets}.

\begin{table}[!htb]
    \centering
    \caption{Dataset Characteristics for \textsc{FedSZ} Benchmarking}      
    \resizebox{\columnwidth}{!}{
    \begin{tabular}{ccccc}
    \toprule
        \textbf{Dataset} & \textbf{\# of Samples} & \textbf{Input Dimension} & \textbf{Classes} \\
            \midrule
CIFAR-10~\cite{cifar10}& $60,000$ & $32\times32$ & $10$ \\
Fashion-MNIST~\cite{fmnist} & $70,000$ & $28 \times 28$ & $10$ \\
Caltech101~\cite{caltech101} & $9,000$ & $224 \times 224$ & $101$ \\
    \bottomrule
    \hspace{1mm}
    \end{tabular}}
    \label{tab:datasets}
 \vspace{-4mm}
\end{table}
The choice of CIFAR-10, Fashion-MNIST, and Caltech101 serves multiple purposes. CIFAR-10 and Fashion-MNIST are more straightforward datasets, often used for benchmarking. They allow us to assess how well \textsc{FedSZ} performs under less demanding conditions. Caltech101, with its higher image pixel count and greater number of classes, is a more challenging image classification problem, testing that the error introduced by our compression doesn't just work with "toy" tasks. These datasets provide a comprehensive test bed for evaluating \textsc{FedSZ} across various scenarios. We perform our testing using FedAvg for four clients and fifty communication rounds with one epoch per client per round.

% \begin{table*}[!ht]
%     \centering
%     \caption{Compression Ratios for \textsc{FedSZ} for Various Models and Datasets}    
%     \vspace{-2mm}
%     \begin{tabular}{ccccccccccccc}
%         \toprule
%         Dataset & \multicolumn{4}{c}{CIFAR-10} & \multicolumn{4}{c}{Caltech101} & \multicolumn{4}{c}
%         {Fashion-MNIST} \\
%         \toprule
%         REL Error Bound & $10^{-1}$ & $10^{-2}$ & $10^{-3}$ & $10^{-4}$ & $10^{-1}$ & $10^{-2}$ & $10^{-3}$ & $10^{-4}$ & $10^{-1}$ & $10^{-2}$ & $10^{-3}$ & $10^{-4}$ \\
%         \midrule
%         AlexNet     & 54.54 & 12.61 & 5.54 & 3.52 & 51.19 & 9.62 & 4.94 & 3.27 & 186.42 & 12.01 & 5.69 & 3.57 \\
%         MobileNetV2 & 11.07 & 5.39 & 3.23 & 1.94 & 10.62 & 5.26 & 3.18 & 1.93 & 11.70  & 5.55  & 3.27 & 1.98 \\
%         ResNet50    & 20.21 & 7.02 & 4.04 & 2.73 & 17.81 & 6.68 & 3.90 & 2.66 & 22.76  & 7.11  & 4.12 & 2.74 \\
%         \bottomrule
%     \end{tabular}    \label{tab:compression_ratios}
%     \vspace{-4mm}
% \end{table*}
\begin{table*}[!htb]
    \centering
    \caption{Compression Ratios for \textsc{FedSZ} for Various Models and Datasets}    
    \vspace{-2mm}
    \begin{tabular}{ccccccccccccc}
        \toprule
        Dataset & \multicolumn{4}{c}{CIFAR-10} & \multicolumn{4}{c}{Caltech101} & \multicolumn{4}{c}
        {Fashion-MNIST} \\
        \toprule
        REL Error Bound & $10^{-1}$ & $10^{-2}$ & $10^{-3}$ & $10^{-4}$ & $10^{-1}$ & $10^{-2}$ & $10^{-3}$ & $10^{-4}$ & $10^{-1}$ & $10^{-2}$ & $10^{-3}$ & $10^{-4}$ \\
        \midrule
        AlexNet     & $54.54$ & $12.61$ & $5.54$ & $3.52$ & $51.19$ & $9.62$ & $4.94$ & $3.27$ & $186.42$ & $12.01$ & $5.69$ & $3.57$ \\
        MobileNetV2 & $11.07$ & $5.39$ & $3.23$ & $1.94$ & $10.62$ & $5.26$ & $3.18$ & $1.93$ & $11.70$  & $5.55$  & $3.27$ & $1.98$ \\
        ResNet50    & $20.21$ & $7.02$ & $4.04$ & $2.73$ & $17.81$ & $6.68$ & $3.90$ & $2.66$ & $22.76$  & $7.11$  & $4.12$ & $2.74$ \\
        \bottomrule
    \end{tabular}    
    \label{tab:compression_ratios}
    \vspace{-4mm}
\end{table*}

\subsection{Simulating Network Bandwidth}\label{sec:network}

To evaluate \textsc{FedSZ}'s communication reduction in low bandwidth network conditions, we emulate different bandwidths by introducing delays into MPI when sending data between processes. We make a note that part of \textsc{FedSZ}'s evaluation is on a cluster where we use MPI in APPFL for simulating multiple FL clients, therefore this simulation is from our cluster deployment and not the Raspberry Pi 5. 

The goal of this method is to demonstrate how \textsc{FedSZ} would perform with low bandwidth, which is common in real-world FL deployments. We are able to simulate this effect by first measuring the process-to-process MPI bandwidth. Then, we use \texttt{sleep} to wait for the length of time it would take to transmit our compressed client update proportional to our desired bandwidth. By incorporating these controlled sleep delays into the communication protocol, we simulate the low-bandwidth environments typical of edge devices (e.g., $10\si{Mbps}$~\cite{edge-bandwidth}) compared to the data center setting which can approach $10\si{Gbps}$. This strategy allows us to demonstrate the communication runtime reduction offered by \textsc{FedSZ}'s compression techniques.

\section{Performance Evaluation}
Having outlined the guiding design for \textsc{FedSZ} and our evaluation methodology, we present results demonstrating the impact of \textsc{FedSZ} on the inference accuracy, runtime, and communication time of FL and the scalability of FL with \textsc{FedSZ} through strong and weak scaling. In our evaluation, we prototype rounds of learning on the Argonne Swing cluster to accelerate learning tasks and measure compression runtime results from a Raspberry Pi 5. Therefore, for our experiments showing the learning and accuracy capabilities of \textsc{FedSZ} during training, we use the system described in Section \ref{subsec:lossy-comp-methods} and for runtime and compression benchmarking we use data collected from a Raspberry Pi 5.

\subsection{Compression Impacts on Inference Accuracy}

An important consideration in our study is the trade-off between compression ratio and model accuracy when utilizing \textsc{FedSZ} with \texttt{SZ2} and \texttt{blosc-lz} compressors. Therefore, we test the impact of varying \texttt{SZ2}'s relative (REL) error bounds from $10^{-5}$ to $10^{-1}$. Our findings, illustrated in Figure \ref{fig:error-overheads}, reveal a clear threshold at $10^{-2}$ beyond which model accuracy sharply declines due to the magnitude of the error introduced by the compressor. This decline indicates a boundary where a loss in model utility offsets the benefits of higher compression (and thus reduced communication costs). However, for error bounds less than or equal to $10^{-2}$, we observe that the inference accuracy remains within 0.5\% of the uncompressed model's performance. At some error bounds, there is a deviation from this 0.5\%. However, this is due to the natural variability of training and validation. This is a crucial observation, underscoring that we can compress FL client updates without affecting a model's efficacy. 

Moreover, the compression ratios presented in Table \ref{tab:compression_ratios} further corroborate our recommendation of $10^{-2}$ as an optimal REL. For instance, at a REL of $10^{-2}$, AlexNet achieved a compression ratio of 12.61$\times$ on CIFAR-10, a significant gain without damaging model accuracy. We observe similar trends for MobileNetV2 and ResNet50 across all datasets. The consistency in these trends across various model architectures and datasets suggests that a REL of $10^{-2}$ could be a generalized optimal setting in a wide range of FL applications. Similarly, as shown in Figure \ref{fig:lossy-accuracy}, we see that \textsc{FedSZ} did not significantly impact the rate of convergence for models. The shading in this plot represents the standard deviation of accuracy over all the explored relative error bounds $<10^{-2}$, revealing that there is no impact on convergence for error bounds less than this value.

Taking into account Eqn.~\ref{eqn:second-formulation}, we further the argument that choosing of error bound is a critical factor in achieving an effective balance between compression efficiency and model accuracy. Therefore, based on our results, we recommend a relative error bound of $10^{-2}$ as it offers communication runtime reduction without compromising model performance, making it well-suited for FL optimizing client-server communications.

% \begin{figure}[H]
%     \centering
%     \includegraphics[width=0.9\columnwidth]{Caltech101-error.pdf}
%     \caption{Caltech101 Inference Accuracy over Different Error Bounds}
%     \label{fig:caltech-error}
% \end{figure}

% \begin{figure}[H]
%     \centering
%     \includegraphics[width=0.9\columnwidth]{CIFAR10-error.pdf}
%     \caption{CIFAR10 Inference Accuracy over Different Error Bounds}
%     \label{fig:cifar-error}
% \end{figure}

% \begin{figure}[H]
%     \centering
%     \includegraphics[width=0.9\columnwidth]{FMNIST-error.pdf}
%     \caption{FMNIST Inference Accuracy over Different Error Bounds}
%     \label{fig:fmnist-error}
% \end{figure}

\subsection{Time Overhead of \textsc{FedSZ} and Network Gains}\label{sec:time-gains}
\begin{figure*}[!ht]
  \centering
  \vspace{5mm}
\subfigure[{Caltech101}\hspace{-5mm}]
{
\label{fig:caltech-time}
\raisebox{-1cm}{\includegraphics[scale=0.3]{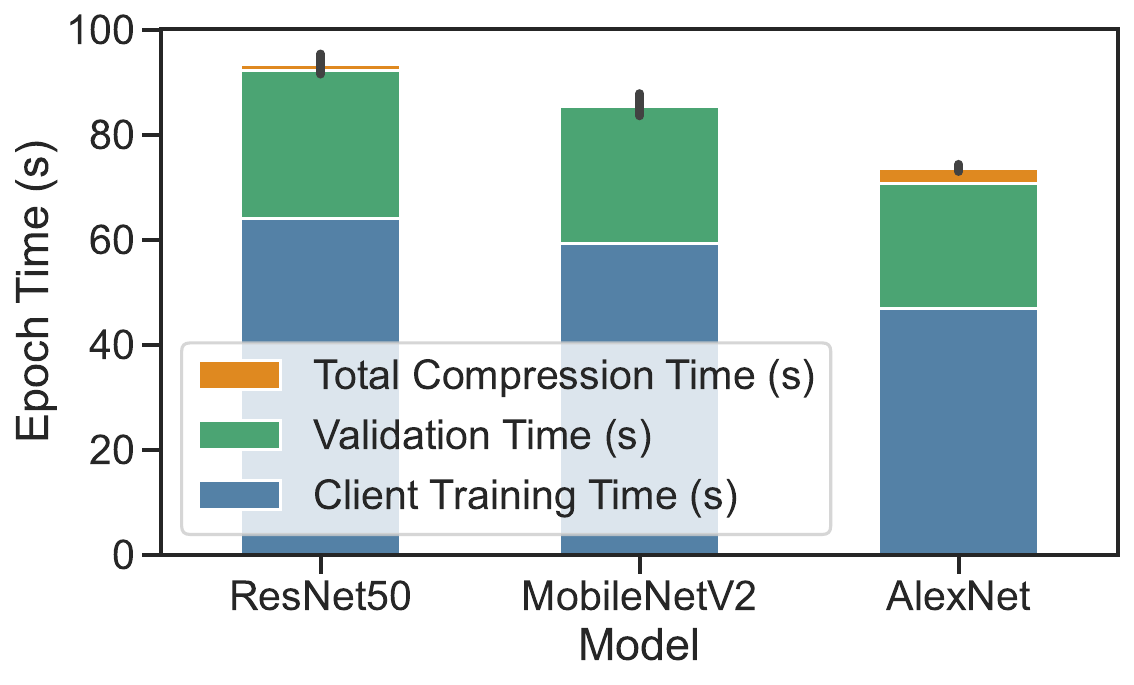}}
}
\hspace{-4mm}
\subfigure[{CIFAR-10}\hspace{-3mm}]
{
\label{fig:cifar-time}
\raisebox{-1cm}{\includegraphics[scale=0.3]{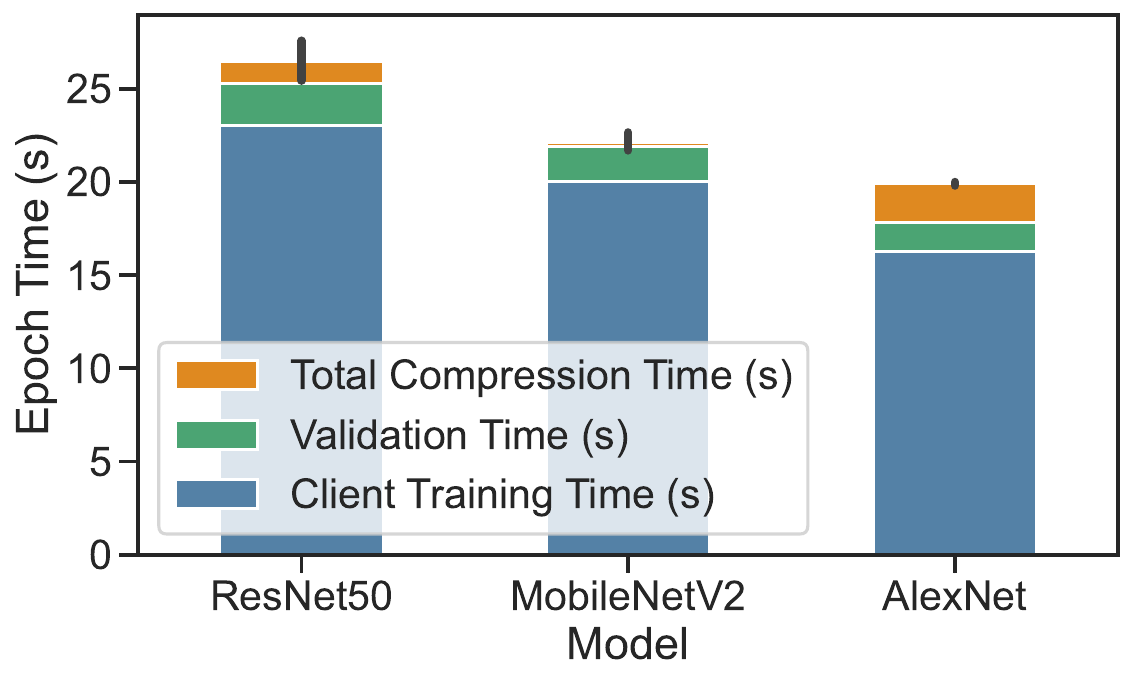}}
}
\hspace{-4mm}
\subfigure[{Fashion-MNIST}\hspace{-5.5mm}]
{
\label{fig:fmnist-time}
\raisebox{-1cm}{\includegraphics[scale=0.3]{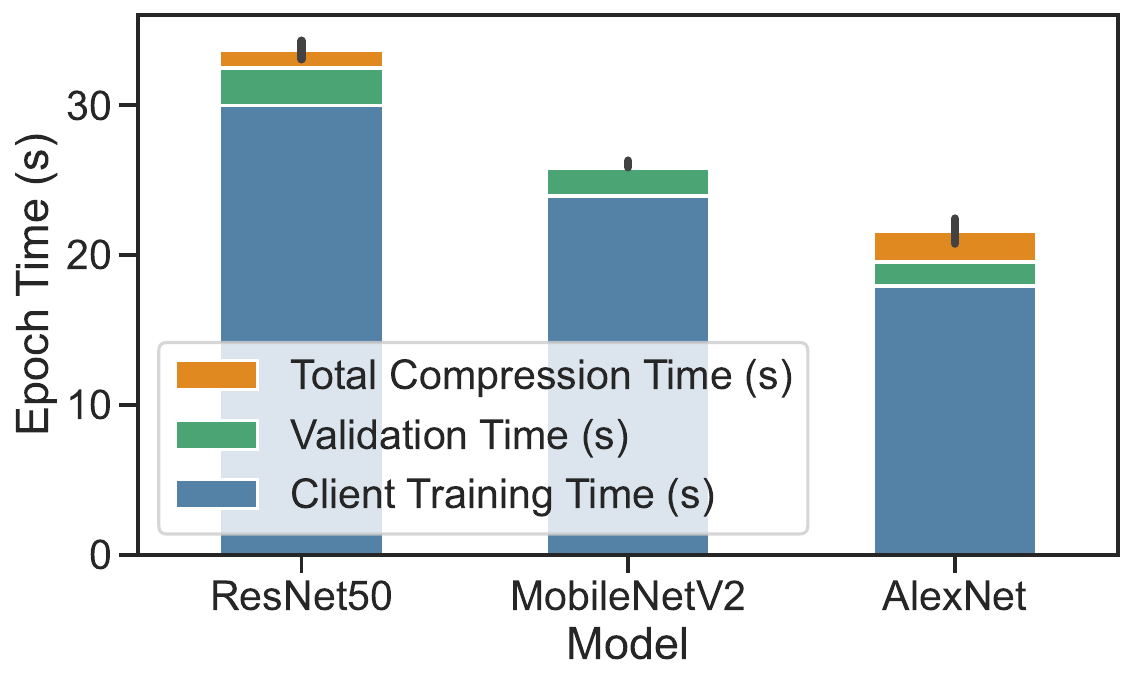}}
}
\vspace{-3mm}
  \caption{Client Runtime per Epoch Breakdown including \textsc{FedSZ} Compression}
  \label{fig:time-overheads}
  \vspace{-3mm}
\end{figure*}

\subsubsection{Cluster Setting} A primary concern of \textsc{FedSZ} is whether the runtime overhead introduced by compressing and decompressing model updates outweighs the gains in reduced communication runtime (see Eqn~\ref{eqn:comm-ineq}); however, we find that the advantages of \textsc{FedSZ} in communication runtime reduction outweighs the introduced compression overhead. For a cluster setting, Figure \ref{fig:time-overheads} shows the mean runtime per epoch across various models and datasets when using \textsc{FedSZ} with an error bound of $10^{-2}$. The compression process takes additional runtime, at its most extreme case with AlexNet on CIFAR-10, the overhead was 3.66 seconds or 17\% of the total runtime. In the rest of the cases, the wall-clock overhead was $<$$12.5\%$, and an average of $4.7\%$ of the client's total epoch time. These times, however, are relatively minor when evaluated in the context of the overall communication time savings that compression offers, particularly at larger error bounds.

\begin{figure*}[!htb]
    \centering
    \captionsetup{justification=centering}
\subfigure[{AlexNet}\hspace{-6mm}]
{
\label{fig:alexnet-mpi}
\raisebox{-1cm}{\includegraphics[scale=0.32]{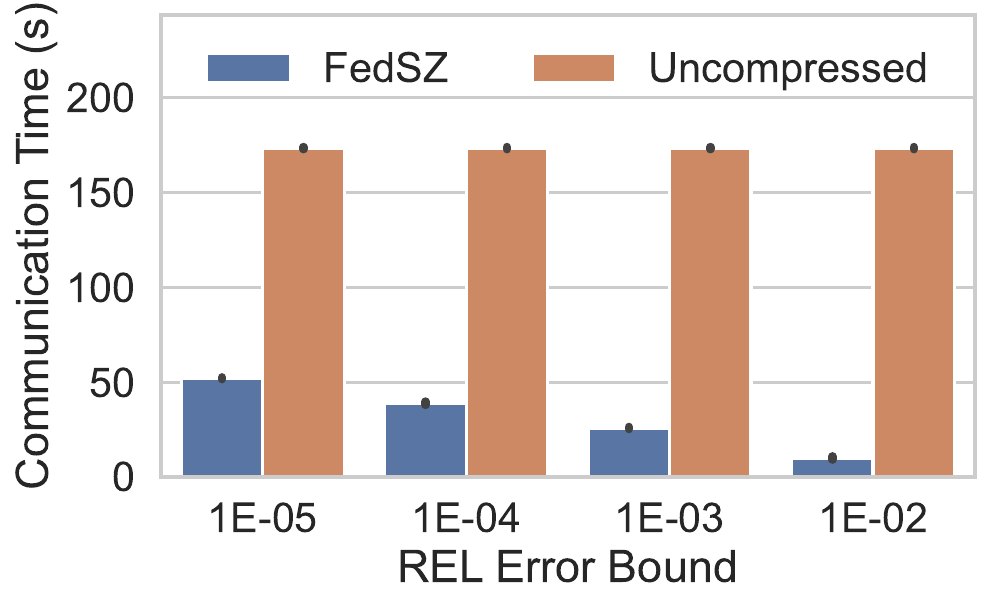}}
}
\hspace{-4mm}
\subfigure[{MobileNetV2}\hspace{-5mm}]
{
\label{fig:mobilenet-mpi}
\raisebox{-1cm}{\includegraphics[scale=0.32]{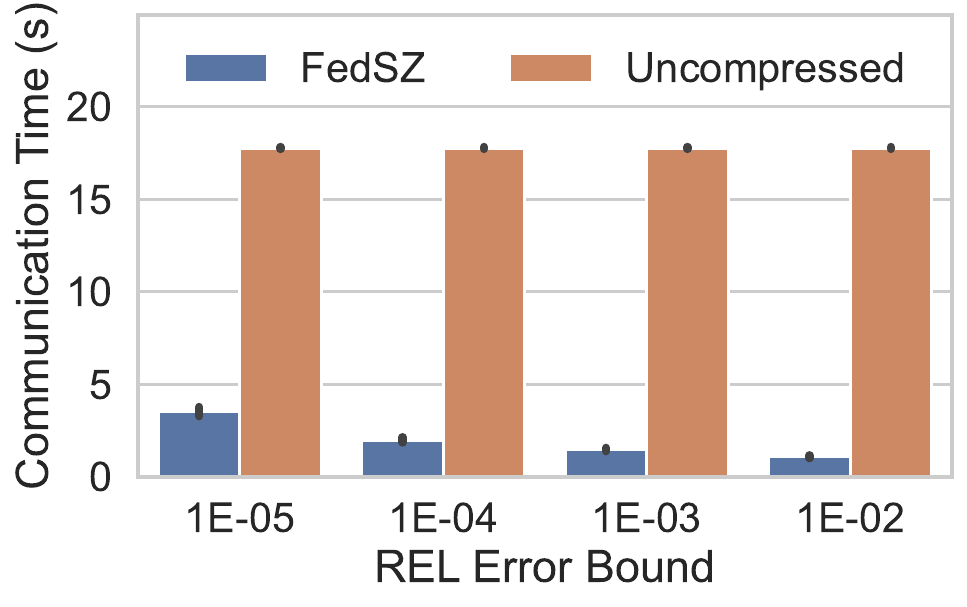}}
}
\hspace{-3mm}
\subfigure[{ResNet50}\hspace{-6mm}]
{
\label{fig:resnet-mpi}
\raisebox{-1cm}{\includegraphics[scale=0.32]{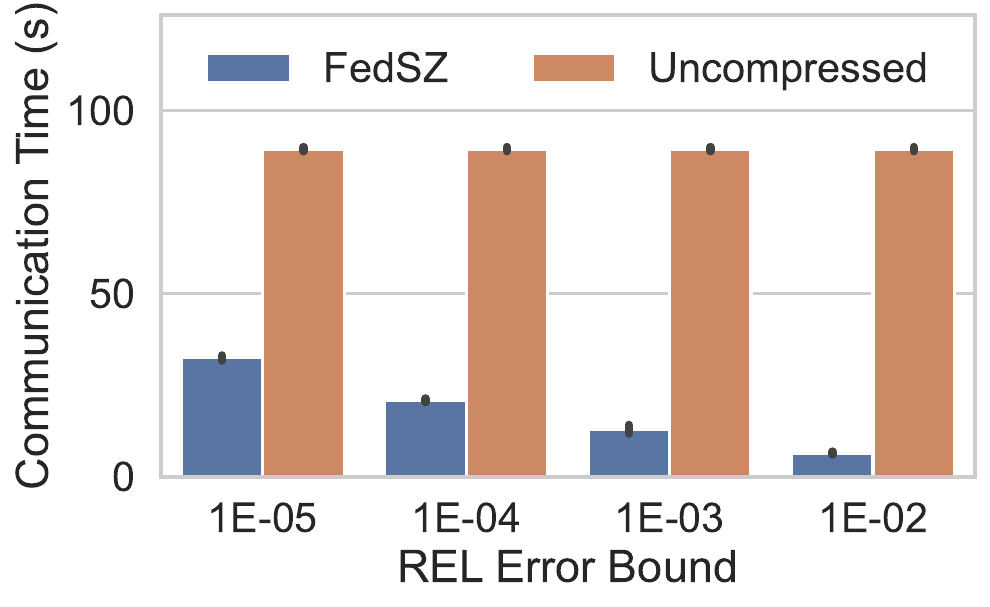}}
}\vspace{-3mm}
    \caption{Total Communication Time for Models over Different REL Error Bounds on $10\si{Mbps}$ Network}
    \label{fig:mpi-timing}
    \vspace{-5mm}
\end{figure*}

Figure \ref{fig:mpi-timing} shows the simulated communication time of a client update to the server, including compression and decompression, on a network of $10\si{Mbps}$; when we compress at any error bound, we decrease the communication time by an order of magnitude from uncompressed data transmission. For example, at an error bound of $10^{-2}$ on a simulated $10\si{Mbps}$network, AlexNet experienced a $109.87\si{s}$ or $13.26\times$ reduction in communication time. Similarly, reductions for MobileNetV2 and ResNet50 were by $12.23\%$ and $9.74\%$, respectively. This means that the server receives updates faster and can begin aggregating the results to begin the next communication round, and the client is using less runtime on I/O as previously.

\subsubsection{Edge Setting} The merit of employing \textsc{FedSZ} is clear when examining the communication time savings in a limited network bandwidth setting. Since FL is typically decentralized learning on geographically distributed edge devices, we include benchmarks of the compression overhead when using \textsc{FedSZ} on a Raspberry Pi 5 in Table \ref{table:multi_model_comparison}. Some quantities, such as accuracy and compression ratio, are hardware independent, so a key metric then is how long compression takes on a certain system. The importance of this point is made clear in Eqn. \ref{eqn:comm-ineq}, where we see that compression for communications is advantageous with a low compression runtime and high compression ratio. We should note that this is irrespective of and has no effect on training time, as \textsc{FedSZ} is post-training. 

\begin{figure}[H]
    \centering
    \captionsetup{justification=centering}
    \includegraphics[width=\columnwidth]{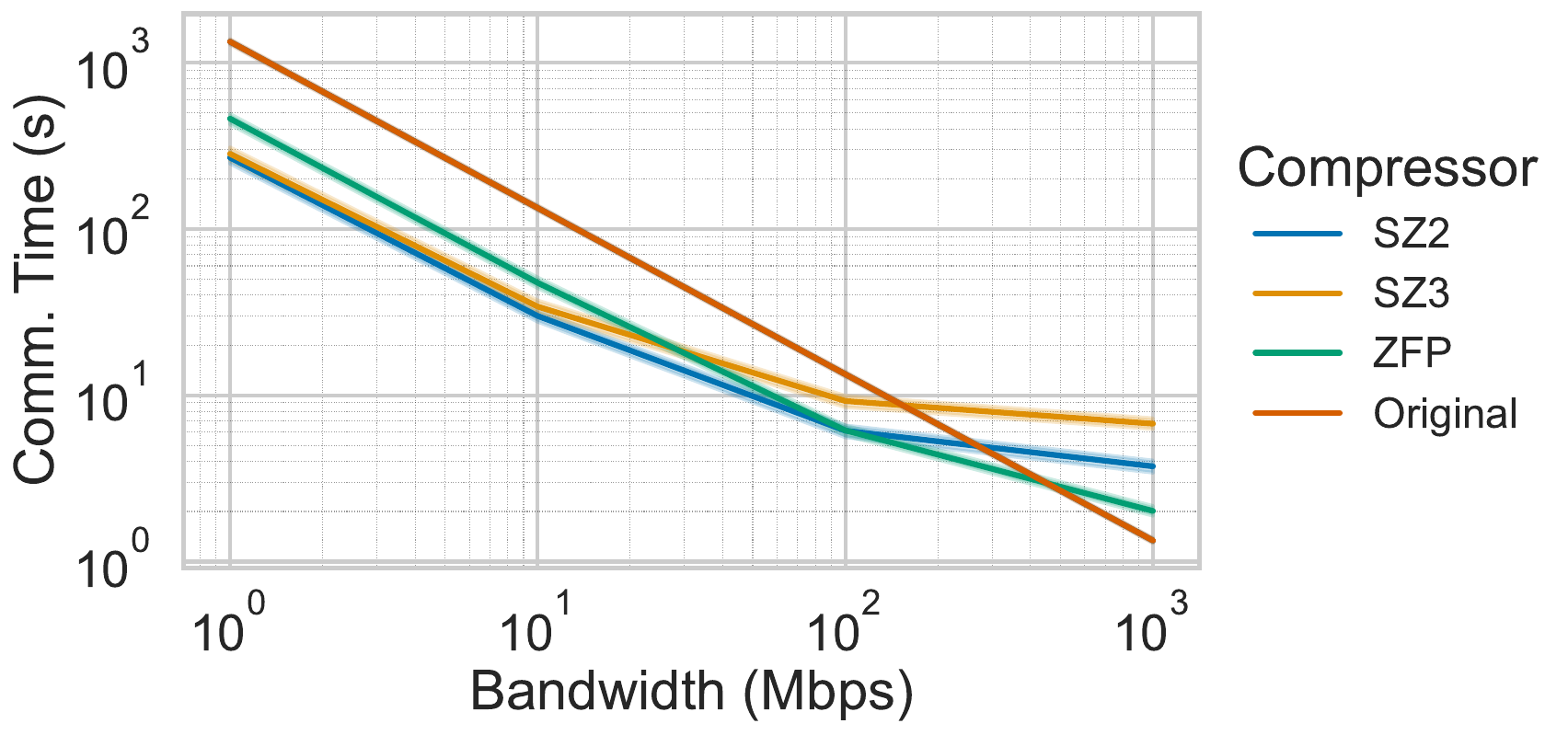}
    \caption{Communication Time for Transmitting AlexNet over Variable Network from Raspberry Pi 5}
    \label{fig:transmission}
    \vspace{-3.5mm}
\end{figure}

An interesting consequence of Equation \ref{eqn:comm-ineq} is it reveals whether you should compress for a given network bandwidth. Using the runtime overhead and compression ratios from Tables \ref{table:multi_model_comparison} and \ref{tab:compression_ratios}, respectively, we can calculate the communication runtime for a spectrum of bandwidths to generate Figure \ref{fig:transmission}.
Here we notice that until approximately $500\si{Mbps}$ compression is optimal, and until $100\si{Mbps}$ \texttt{SZ2} is most optimal, before the runtime overhead of compression takes outweighs the gains in compression ratio. Therefore on a $<500\si{Mbps}$ wide area network with edge device compression latency, it is more runtime efficient to compress before communication. The gains in communication reduction before this threshold are significant before this point, as a client can save an order of magnitude of runtime when transmitting a model update to the server.

\subsection{Scalability of \textsc{FedSZ}}

To test whether \textsc{FedSZ} scales effectively with the number of clients in a system we evaluate scalability up to 128 CPU cores on Argonne's Swing cluster, increasing the CPU core count in powers of 2 while simulating a network bandwidth of $10\si{Mbps}$ using the method described in Section~\ref{sec:network}. 

\begin{figure}[!htb]
    \centering
    \captionsetup{justification=centering}
    \hspace{-4mm}
\subfigure[{Weak Scaling Results}\hspace{-4mm}]
{
\label{fig:weak-scaling}
\raisebox{-1cm}{\includegraphics[scale=0.33]{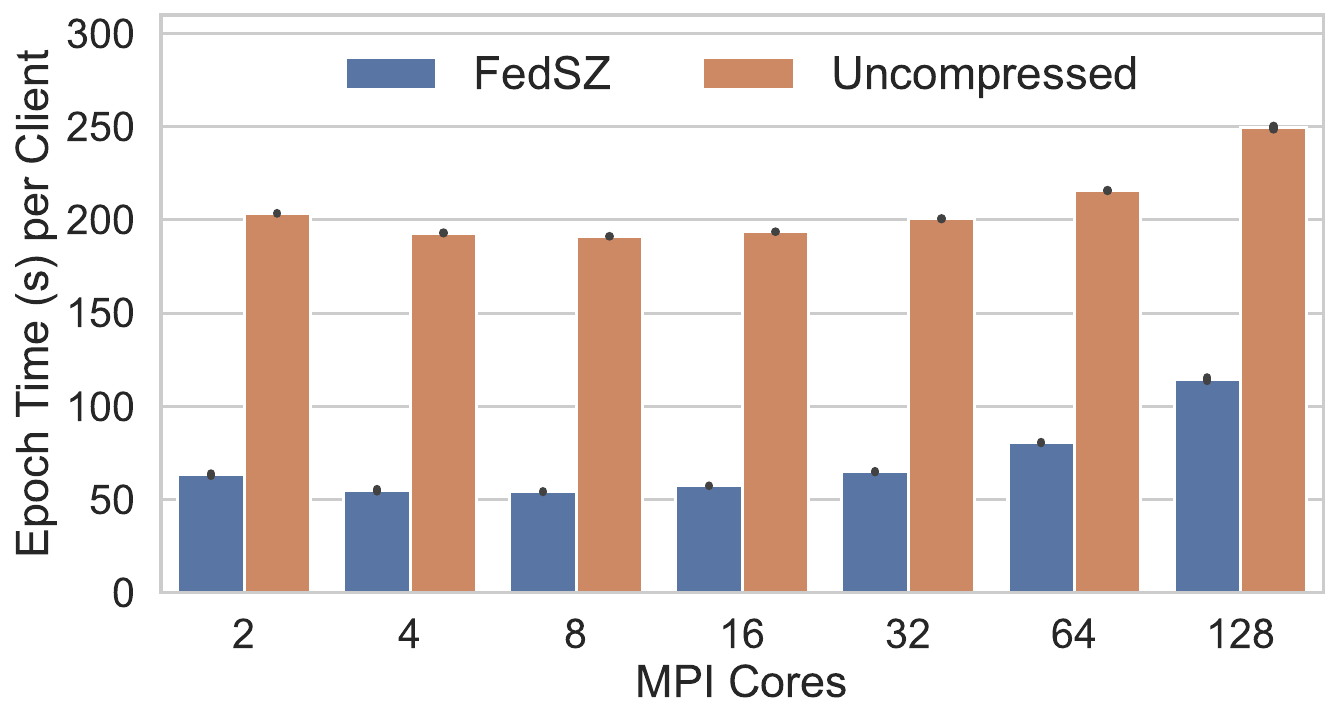}}
}
\hspace{-5mm}
 \subfigure[{Strong Scaling Results}\hspace{-4mm}]
{
\label{fig:strong-scaling}
\raisebox{-1cm}{\includegraphics[scale=0.33]{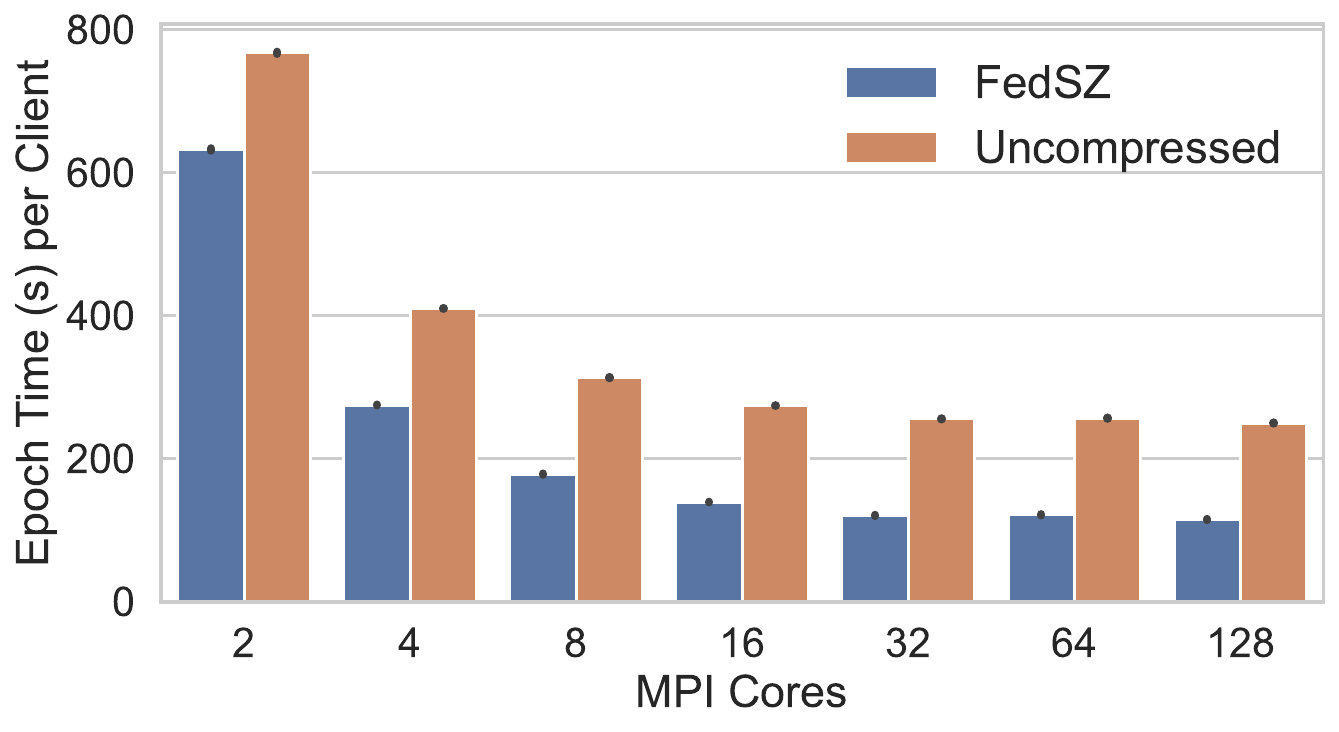}}
}   
    \caption{Scaling Results for Training MobileNet-V2 on CIFAR-10 with and without \textsc{FedSZ}}
    \label{fig:scaling-results}
    %\vspace{-5mm}
\end{figure}

For our weak scaling test, we assign one client per core while increasing the total number of CPU cores. \textsc{FedSZ} demonstrates effective weak scaling, as evidenced by the near-linear increase in communication time as the number of cores and clients increased, shown in Figure~\ref{fig:weak-scaling}. In our evaluation, \textsc{FedSZ} exhibits a recalculated weak scaling speedup ranging from 0.36 at 128 MPI processes to a peak of 1.64 with 8 MPI processes, indicating moderate adaptability to an increasing client count. 

For our strong scaling test, we kept 127 clients constant while increasing the total number of CPU cores. \textsc{FedSZ} shows robust, strong scaling characteristics as shown in Figure \ref{fig:strong-scaling}. In this experiment, the framework achieves a recalculated speedup as high as 7.51 at 128 MPI processes, demonstrating effective utilization of additional computational resources for a fixed number of clients.

Simulating a constrained network environment of $10\si{Mbps}$ demonstrates the efficacy of \textsc{FedSZ}'s compression algorithms. Under these conditions, the framework is able to maintain the model's accuracy while significantly reducing the communication overhead. The advantage of using compression becomes increasingly evident as the number of cores (and thus, the number of clients) increases. This suggests that \textsc{FedSZ}'s compression techniques can offer substantial benefits in low-bandwidth scenarios, where communication costs are often high.

\subsection{Potential Differential Privacy of Lossy Compression}

\begin{figure}[H]
    \centering
    \includegraphics[width=\columnwidth]{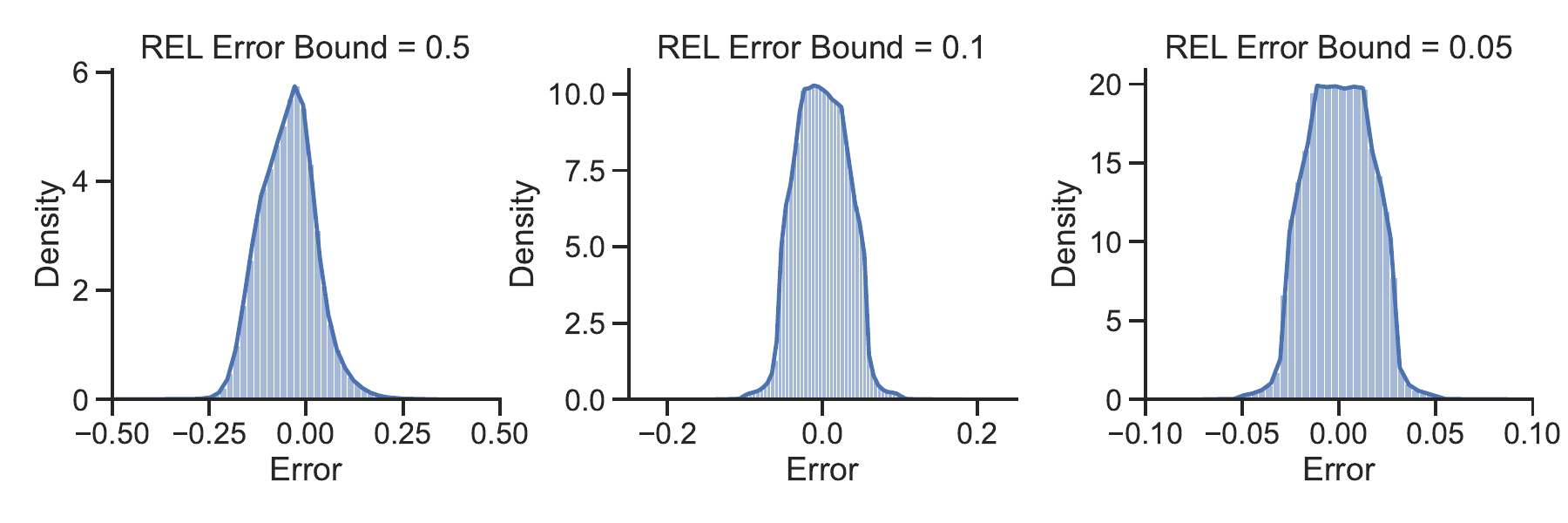}
    \vspace{-3mm}\caption{Distribution of Errors for Different Error Bounds}
    \label{fig:diff-privacy}
\end{figure}

An intriguing aspect of \textsc{FedSZ} is that lossy compression adds some noise to the data after decompression, a phenomenon that could potentially introduce to differential privacy (DP). A user can find the distribution of errors by taking the pairwise difference of the original weights and the decompressed weights. Our experiments have shown that error distributions of the communicated model parameters exhibit characteristics similar to a Laplacian distribution, as evidenced by the histograms in Figures \ref{fig:diff-privacy}. 
DP can work by adding noise with a certain random distribution to data before communication (e.g. Laplacian distributed noise) to ensure that an individual's data cannot be easily identified from aggregate statistics~\cite{dwork2006calibrating}. It is worth noting that the resemblance to Laplacian noise does not guarantee DP, as the formal guarantee depends on specific mathematical conditions related to the data sensitivity and the noise scale. However, the observed characteristics in our study make this an avenue worth exploring. Chen et al.~\cite{chen2024privacy} corroborate these claims with their findings that types of compression can introduce DP, further meriting exploration into whether EBLCs possess the same properties.  Given the critical importance of data privacy in FL, this warrants further investigation and could add another layer of utility to \textsc{FedSZ}'s already promising performance.

\section{Discussion and Future Work}\label{sec:con}

\subsection{Takeaways and Addressed Questions}

\begin{itemize}
    \item \textbf{How to integrate lossy compressors in FL communication?}
        \begin{itemize}
            \item We develop \textsc{FedSZ}, a robust algorithm that seamlessly integrates lossy compression with \texttt{SZ2} and \texttt{blosc-lz} into the FL client-server communication pipeline. Our implementation supports various model architectures and datasets, offering a modular approach for FL applications.
        \end{itemize}
    \item \textbf{What EBLC would perform best given the task of compressing FL client updates?}
    \begin{itemize}
        \item Through empirical evidence, we find that \texttt{SZ2} consistently outperforms other compressors with respect to compression ratio while maintaining uncompressed model accuracy. It offers a balance between compression efficiency and minimal impact on model performance.
    \end{itemize}
    \item \textbf{EBLCs introduce error and time overhead. Is it worth using this as a data reduction strategy?}
    \begin{itemize}
        \item Our results indicate that for relative error bounds up to $10^{-2}$, \textsc{FedSZ} maintained model accuracy within 1\% of the uncompressed model while achieving significant compression ratios. Thus, the minor time overhead of compression is justifiable given the substantial reduction in communication cost. 
        \end{itemize}
\end{itemize}

\subsection{Future Directions}
This study presents a robust compression algorithm for reducing the communication overhead of client-server interactions for FL. Future research directions beyond the scope of this study are to explore (1) how hyperparameter optimization might be tuned to mitigate the accuracy loss introduced by compression, thereby leading to more optimal model performance and (2) how noise might offer DP for communications, a concept crucial for ensuring data privacy in FL. Future studies could examine the relationship between the noise generated by lossy compression and its impact on DP guarantees. These areas all warrant subsequent exploration to increase the merits of including lossy compression in this technology.

\section{Conclusion}
Our study demonstrates that EBLCs can mitigate communication overhead in FL client-server communications without compromising model accuracy. This is particularly significant in scenarios where bandwidth is constrained, as our algorithm achieves remarkable compression ratios and maintains inference accuracy within small relative error bounds. Looking ahead, the potential of \textsc{FedSZ} extends beyond current applications. We envisage its integration into a broader array of FL frameworks and scenarios, potentially enhancing data privacy through DP techniques inspired by the noise characteristics inherent in lossy compression processes. Not only this, but \textsc{FedSZ} as a last-step in the communication pipeline will work effectively with other existing compression techniques for FL such as gradient sparsification, pruning, and quantization. Moreover, the open-source availability of \textsc{FedSZ} within the APPFL framework promises to allow for usage beyond the scope of this project. 

\section*{Acknowledgment}
This research was supported by the U.S. Department of Energy, Office of Science, Advanced Scientific Computing Research (ASCR), under contract DE-AC02-06CH11357, and supported by the National Science Foundation (NSF) under Grant OAC-2003709, OAC-2104023, and OAC-2311875. We acknowledge the computing resources provided on Swing (operated by Laboratory Computing Resource Center at Argonne). During the time of this work GW was supported by a Churchill Scholarship.

\bibliographystyle{ieeetr}
\bibliography{main.bib}

\end{document}